\documentclass[acmlarge]{acmart}
 \pdfoutput=1 
\usepackage{booktabs} 
\setcopyright{rightsretained}
\usepackage{graphicx}
\usepackage{subfigure}
\usepackage{amsfonts}

\usepackage{amssymb}
\usepackage{color}
\usepackage{amsmath}
\usepackage{balance}

\usepackage{multirow}

\usepackage[utf8]{inputenc}

\newcommand{\sys}{P2PLocate}

\begin{document}

\setcopyright{acmcopyright}
\acmJournal{IMWUT}
\acmYear{2020} 
\acmVolume{4}
\acmNumber{3} 
\acmArticle{105} 
\acmMonth{9} 
\acmPrice{15.00}
\acmDOI{10.1145/3411833}

\title{Peer-to-Peer Localization for Single-Antenna Devices} 

%anonymous authors
%\author{Anonymous Author(s)}

%authors
\author{Xianan Zhang}
\email{xiananzhang@hust.edu.cn}
\affiliation{%
	\institution{Huazhong University of Science and Technology}
	\streetaddress{1037 Luoyu Road}
	\city{Wuhan}
	\state{Hubei}
	\country{China}}
\author{Wei Wang}
\authornote{This is the corresponding author}
\email{weiwangw@hust.edu.cn}
\affiliation{%
	\institution{Huazhong University of Science and Technology}
	\streetaddress{1037 Luoyu Road}
	\city{Wuhan}
	\state{Hubei}
	\country{China}}
\author{Xuedou Xiao}
\email{xuedouxiao@hust.edu.cn}
\affiliation{%
	\institution{Huazhong University of Science and Technology}
	\streetaddress{1037 Luoyu Road}
	\city{Wuhan}
	\state{Hubei}
	\country{China}}
\author{Hang Yang}
\email{hangyang@hust.edu.cn}
\affiliation{%
	\institution{Huazhong University of Science and Technology}
	\streetaddress{1037 Luoyu Road}
	\city{Wuhan}
	\state{Hubei}
	\country{China}}

\author{Xinyu Zhang}
%\authornote{This is the corresponding author}
\email{xyzhang@ucsd.edu}
\affiliation{%
	%\institution{Department of Electrical and Computer Engineering, University of California San Diego}
	\institution{University of California San Diego}
	%\streetaddress{Clear Water Bay, Kowloon}
	\city{San Diego}
	\state{California}
	\country{USA}}
\author{Tao Jiang}
\email{taojiang@hust.edu.cn}
%\author{XX}
%%\authornote{This is the corresponding author}
%\email{XX}
\affiliation{%
	\institution{Huazhong University of Science and Technology}
	\streetaddress{1037 Luoyu Road}
	\city{Wuhan}
	\state{Hubei}
	\country{China}}

\begin{abstract}
	%backup0213
	Some important indoor localization applications, such as localizing a lost kid
	in a shopping mall, call for a new peer-to-peer localization technique that
	can localize an individual's smartphone or wearables by directly using
	another's on-body devices in unknown indoor environments. However, current
	localization solutions either require pre-deployed infrastructures or multiple 
	antennas in both transceivers, impending their wide-scale application. 
	In this paper, we present \sys, a peer-to-peer
	localization system that enables a single-antenna device co-located with a
	batteryless backscatter tag to localize another single-antenna device with
	decimeter-level accuracy. \sys\ leverages the multipath variations
	intentionally created by an on-body backscatter tag, coupled with spatial
	information offered by user movements, to accomplish this objective without
	relying on any pre-deployed infrastructures or pre-training. \sys\
	incorporates novel algorithms to address two major challenges: (i)
	interference with strong direct-path signal while extracting multipath
	variations, and (ii) lack of direction information while using single-antenna
	transceivers. We implement \sys\ on commercial off-the-shelf Google Nexus 6p,
	Intel 5300 WiFi card, and Raspberry Pi B4. Real-world experiments reveal that \sys\ can localize both static and mobile targets with a median accuracy of 0.88~m.	
	
	%Some appealing indoor localization applications, such as localizing a lost kid in a shopping mall, call for a new peer-to-peer localization technique that can localize an individual's smartphone or wearables using only another's on-body devices in unknown indoor environments. However, current localization solutions either require pre-deployed infrastructures or the assistance of Inertial Measurement Unit (IMU) which suffers from significant accumulative errors. In this paper, we present \sys, a peer-to-peer localization system that enables a single-antenna device to localize another single-antenna device with decimeter-level accuracy. \sys\ includes a small-formed and battery-less backscatter tag, which either reflects or absorbs signals. The state of the backscatter tag is utilized, along with spatial information offered by the user movement, to achieve accurate peer-to-peer localization without relying on any pre-deployed infrastructures or pre-training. \sys\ also presents novel algorithms that address two aspects of challenges: (i) interference with strong direct-path signal while extracting multipath variations; (ii) lack of direction information when employing just a single antenna in both transceivers. We implement \sys\ on off-the-shelf commodity devices including Intel 5300 WiFi card, Raspberry Pi B4 and Google Nexus 6p. Real-world performance reveals that \sys\ can localize both static and mobile targets with a median accuracy of 0.88~m.   
\end{abstract}


%\ccsdesc[500]{Security and privacy~Security services}
%\ccsdesc[500]{Security and privacy~Mobile and wireless security}


\maketitle
\renewcommand{\shortauthors}{Zhang et al.}

\section{Introduction}\label{sec:Introduction}

%\begin{figure}
%	\centering
%	\includegraphics[width=0.8\linewidth]{figures/Fig1v2}
%	\caption{Motivation example of \sys: finding a lost kid in a shopping mall.}
%	\label{fig:1}
%\end{figure}

%Peer-to-peer localization represents a highly underexplored mobile application scenario but is highly desired in many practical and critical use cases. For example, consider the scenario illustrated in Fig.~\ref{fig:1} where a parent is looking for her lost kid in a shopping mall. According to~\cite{child}, there are nearly one million lost kids worldwide each year. Current solutions rely mainly on law enforcement or store employees to search for the kid, which is time-consuming. Being able to accurately localize the lost kid using RF technologies allows parents to be notified by their everyday-carry smartphones and promptly respond when their kids equipped with smartwatches go out of sight. This demands an RF localization system that (i) works in a plug-and-play manner without pre-training or pre-deployed infrastructures, and (ii) achieves decimeter-level accuracy with minimal antennas that are affordable to on-body mobile devices. Such technology can be applied seamlessly in daily life and is also indispensable for some other emerging applications, e.g., tracking goods in a messy warehouse and relocating medical equipment in an emergency hospital.

Localization service is highly desired in many practical and critical use cases. For example, it can help warehouse keepers find target goods in a messy warehouse, locate medical equipment in an emergency hospital, alert parents when their kids go out of sight in a shopping mall, and remind patients where to find their pill bottles when needed. These use cases call for a new peer-to-peer localization system that should ideally satisfy the following requirements: (i) it should work in a plug-and-play manner without pre-training or pre-deployed infrastructures, as the accurate locations and orientations of pre-deployed infrastructure in warehouses or shopping malls may be unavailable for localization; (ii) it should achieve decimeter-level accuracy with minimal antennas that are affordable to on-body devices, e.g., smartphones carried by warehouse keepers, parents, and patients. More generally, such a technology can enable a variety of applications and be applied seamlessly in our daily lives.

Recent years have witnessed great advances in indoor localization with decimeter-level accuracy through RF technologies~\cite{zhu2013rssi,yin2017peer,teng2013ev,shu2015gradient,ladd2005robotics}. However, different from what we expect to become a ubiquitous service like GPS in outdoors, current techniques suffer from limitations that prohibit their applications in in-situ peer-to-peer localization scenarios: (i) They require either multiple antennas~\cite{ArrayTrack,kotaru2015spotfi} or ultra-wide signal bandwidth~\cite{vasisht2016decimeter,soltanaghaei2018multipath}, which are not available in on-body mobile devices that generally have only a single antenna and limited transmission frequency bandwidth. (ii) They are infrastructure-based~\cite{sen2013avoiding,sugasaki2017robust} and require in advance the accurate deployment locations or/and orientations of the pre-deployed access points (APs) to perform triangulation or trilateration. (iii) Some techniques~\cite{832252,nister2004visual,sen2012you,chen2017taming} require expensive and recurring signal fingerprinting or the assistance of IMUs, which suffers from significant errors caused by environment variations or measurement accumulation. %To break these stalemates, dedicated technologies tailored for conventional single-antenna on-body devices are required.

\begin{figure}
	\centering
	\includegraphics[width=0.8\linewidth]{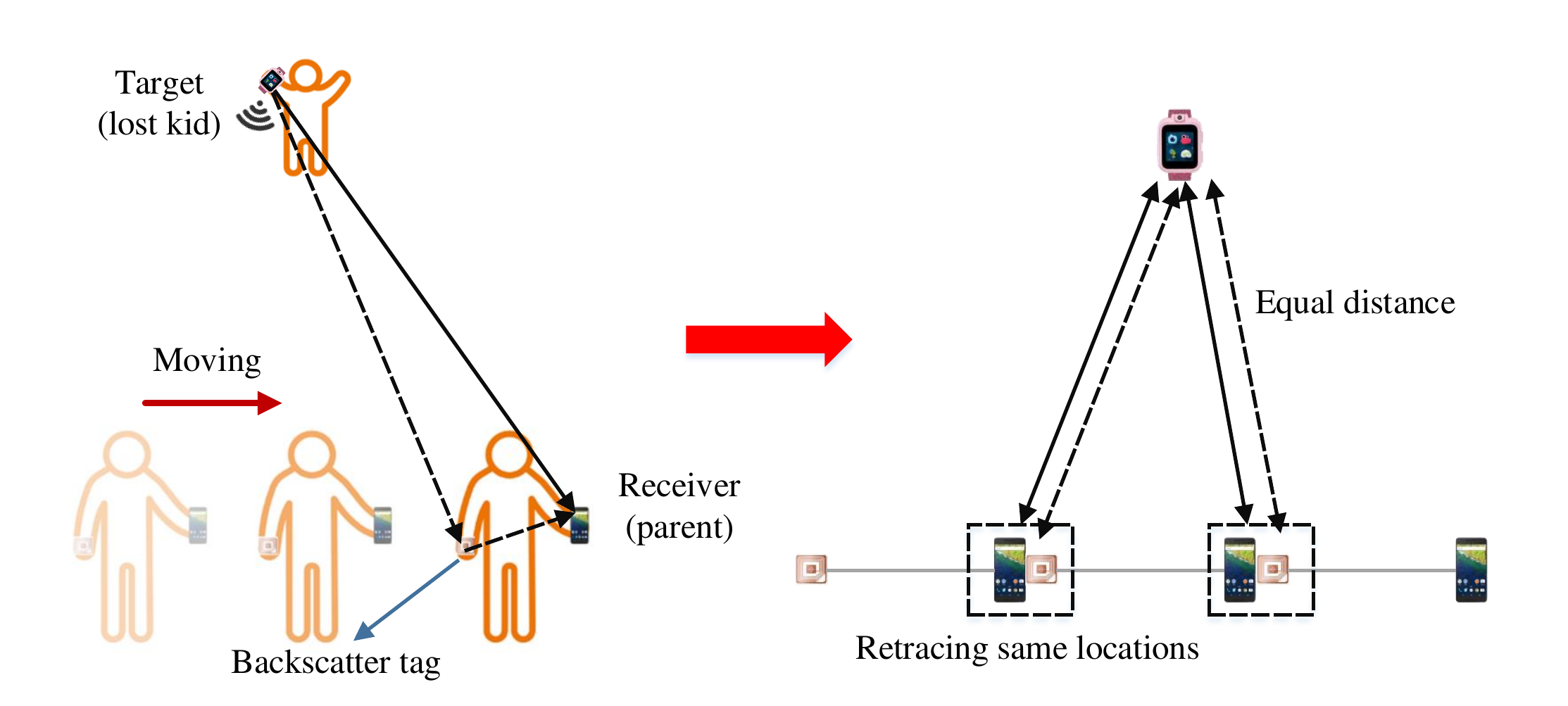}
	\caption{An illustration of direction estimation.}
	\label{fig:keyinsight}
\end{figure}

In this paper, we introduce a peer-to-peer indoor localization system, referred
to as \sys, that enables a single-antenna device (receiver) to localize another
single-antenna device (target) with decimeter-level accuracy. Instead of relying
on multiple antennas or pre-deployed infrastructures, \sys\ employs a single
small-formed, ultra-low-cost, and batteryless backscatter tag. The backscatter
tag communicates by either reflecting or absorbing signals without emitting any power
of its own. In particular, we control the state of the backscatter tag to
change the wireless channel and create multipath variations intentionally. The
multipath variations are further coupled with user movement to pinpoint the
direction of incoming RF signals. Thus, by fusing the direction and range
estimates, the receiver can localize the target. More specifically, \sys\ works
as follows: the target device, which is carried by the lost kid, transmits WiFi
packets. The backscatter tag, attached to the parent’s body, modulates its
information and reflects the WiFi packet. The mobile phone or other wearables
with the commodity WiFi chip, which is carried by the parent, serves as the
receiver to measure the Channel State Information (CSI) of each WiFi packet from
both the remote target device and the backscatter tag and then localizes the
target. In contrast to most prior indoor localization systems, \sys\ can pinpoint the location of any WiFi devices regardless of their antenna numbers, by merely using a single-antenna device and a co-located backscatter tag.

%In this work, a low-cost, small form-factor backscatter tag is embedded into an on-body device, such as a smartwatch or a wristband. Another co-located device that has a commodity WiFi chip like a smartphone serves as a receiver to measure the Channel State Information (CSI) of each WiFi packet from both the remote target device and the backscatter tag and then localizes the target. In contrast to many prior indoor localization systems, \sys\ can pinpoint the location of any WiFi devices regardless of their antenna numbers, by merely using a single-antenna device and a co-located backscatter tag.

\sys\ leverages an unseen opportunity with a single-antenna device and a co-located backscatter tag. We illustrate the intuition in Fig.~\ref{fig:keyinsight}. When the receiver and the co-located backscatter tag move, the backscatter tag may retrace the locations of the receiver. The distance between the target and the backscatter tag would be equal to the distance between the target and the receiver when and only when the backscatter tag arrives at a location traveled by the receiver. Based on this observation, \sys\ can estimate the moving speed of the target relative to the receiver. Thus, by estimating and comparing with the radial speed offered by Doppler effect, \sys\ can identify the direction of the target relative to the receiver.

To translate the above idea into a practical system, however, is nontrivial due to three main challenges. First, the backscatter signal is weak and typically superimposed with a strong direct-path signal as well as signals reflected from furniture, walls and other nearby clutter. Second, the mobility of the receiver is used to identify the direction information of the target, which fails when the target device is also mobile. Third, due to the existence of Carrier Frequency Offset (CFO), the Doppler shift estimated by using traditional approaches only has absolute values without arithmetic signs, which fails the direction estimation.

To address the aforementioned challenges, our system, \sys, introduces three
main innovations that together enable it to achieve decimeter-level accuracy.

\begin{itemize}
	\item{\noindent\textbf{Separating Backscatter CSI.}} \sys's first component
	eliminates the interference from the strong direct-path signal and
	separates robust backscatter CSI. The backscatter CSI can in turn be used to estimate the time of flight (ToF) and Doppler shift. To do so, \sys\ leverages the nature of the backscatter modulation that it modulates information by switching its impedance between the reflective and non-reflective state. \sys\ separates the backscatter CSI by subtracting CSI measurements corresponding to different states while eliminating the effect of significant measurement errors.
	\item{\noindent\textbf{Acquiring full Doppler information and range estimates.}} The accuracy of ToF estimation is limited by the transmission frequency bandwidth. Instead of fully relying on ToF, \sys\ fuses it with Doppler shift to achieve fine-grained range estimation. The key insight is that the resolution of Doppler shift, which depends on the signal observation interval, is much fine-grained than the raw ToF estimates. Further, \sys\ also introduces a two-step algorithm to estimate the full Doppler information, including the absolute value and the arithmetic sign.
	\item{\noindent\textbf{Estimating direction for both static and moving targets.}} To tackle the target mobility which fails the direction estimation, \sys\ carefully analyzes the effect of target mobility on direction estimation and proposes a mobility-resilient direction estimation algorithm. Such a design allows \sys\ to identify the direction of the target no matter it is static or moving.
\end{itemize}

We prototype \sys\ based on commercial off-the-shelf devices, including Google Nexus 6p,
Intel 5300 WiFi card, and Raspberry Pi B4, and a customized backscatter tag
based on FS-Backscatter~\cite{zhang2016enabling}. We conduct both controlled
experiments and field study in two different scenarios, including two different
floors of a laboratory and an office building. The evaluation results reveal that \sys\ achieves an average localization accuracy of 0.88~m, while both transceivers equip with only a single antenna each. The results show approximately a $ 10\times $ accuracy improvement compared to IMU-assisted solution and are comparable to the state-of-the-art localization system that requires multiple antennas and multiple distributed receivers.

\begin{figure}
	\centering
	\includegraphics[width=0.9\linewidth]{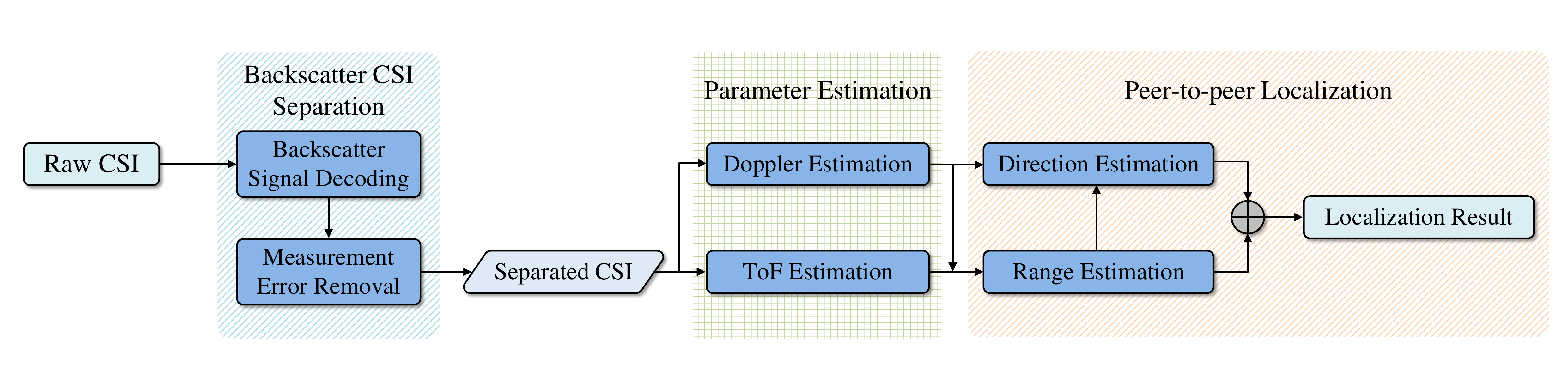}
	\caption{System overview of \sys.}
	\label{fig:system-flow}
\end{figure}

\section{System Overview}\label{sec:overclock}% 1 pp 
\sys\ is a fine-grained peer-to-peer localization system that achieves
decimeter-level accuracy. \sys\ leverages an on-body mobile device and a
co-located backscatter tag to enable an individual to localize another
individual's on-body device with no prior infrastructures or fingerprinting. The
backscatter tag is ultra-low-cost and battery-free and can be easily embedded
into on-body devices. Further, \sys\ only requires a single antenna on both
transceivers and a single-channel measurement. It works both in line-of-sight,
through occlusions, and even when both transceivers move. \sys\ is fully
compatible with commercial off-the-shelf devices equipped with WiFi chips. 

Fig.~\ref{fig:system-flow} shows \sys's three components and its workflow:

\begin{itemize}
	\item \textbf{Backscatter CSI Separation:} \sys\ resolves the backscatter CSI based on the backscatter modulation by using the raw CSI measurements that are exposed by commodity WiFi devices (Section~\ref{sec:design}). 
	%\item Decode and segment the backscatter signals based on the CSI measurements of all subcarriers.
	\item \textbf{Doppler and ToF Estimation:} \sys\ estimates ToF and Doppler shift, including the absolute value and the arithmetic sign corresponding to the direct path from the target device to the receiver and the backscatter tag (Section~\ref{FineDistance}).
	\item \textbf{Peer-to-peer Localization:} \sys\ estimates the range and direction information of the target device relative to the receiver (Section~\ref{DirectionEst}), and then computes the target location by fusing the range and direction estimates. 
\end{itemize}

The next few sections elaborate on the above steps. Some important acronyms and symbols used in this paper are listed in Table~\ref{tab:tab1} and Table~\ref{tab:tab2}, respectively.

\section{Backscatter CSI Separation}\label{sec:design}% 1 pp

In this section, we describe how \sys\ can reliably resolve the backscatter CSI at the receiver side. Recall that the main challenge in backscatter CSI
extraction arises from the strong interference and significant measurement errors. It complicates backscatter signal detection and reduces the reliability of the backscatter CSI.  
% mdTrack && TurboTrack

\subsection{Characterizing Backscatter Communication}\label{backscatter_communication}
Before we describe \sys's backscatter CSI separation algorithm, we start by introducing the backscatter communication technology. In backscatter networking, the backscatter tag communicates by harvesting power from ambient radio frequency (RF) signals, like TV signals, WiFi signals, LoRa signals, etc., which eliminates the requirement of wires and batteries. In \sys, we use the WiFi-based backscatter to assist peer-to-peer localization. Two features of backscatter communication are particularly relevant to the localization problem.

\begin{itemize}
	\item The backscatter tags are battery-free, and can communicate merely by
	leveraging ambient RF signals. It means that once the backscatter tags
	are embedded into the on-body devices, they do not consume any extra
	energy from the devices. Therefore, the on-body devices can afford the utilization of backscatter tags to assist localizing the target. 
	\item Any commodity devices with WiFi chips can receive and decode
	backscatter signals without firmware or hardware modification. The
	complex channel values of the tags can be extracted since many WiFi
	radios have the ability to obtain the CSI per WiFi packet. These channel values can be used for localization. 
	
\end{itemize}

\begin{table}\small
	\centering
	\caption{{Acronyms}}\label{tab:tab1}
	\begin{tabular}{|c|p{3.9cm}|c|p{3.9cm}|}
		\hline
		\textbf{Acronym}     & \textbf{Full name}            & \textbf{Acronym}    & \textbf{Full name} \\ \hline
		CSI                  &   Channel State Information   & ToF                 &  Time of Flight \\  \hline
		CFO                  & Carrier Frequency Offset      & STO                 &  Symbol Time Offset \\  \hline
		SFO                  & Sampling Frequency Offset     & AoA                 &  Angle of Arrival \\  \hline
	\end{tabular}
\end{table}

\begin{table}\small
	\centering
	\caption{{Symbols and notations}}\label{tab:tab2}
	\begin{tabular}{|c|p{4cm}|c|p{4.4cm}|}
		\hline
		\textbf{Symbol}   & \textbf{Definition}             & \textbf{Symbol}     & \textbf{Definition} \\ \hline
		$\textbf{H}_{o}$  & Observed wireless channel at the receiver      & $\textbf{H}_{t}$    &  Wireless channel without backscatter reflections \\  \hline
		$\textbf{H}_{b}$  & Wireless channel corresponding to backscatter reflections          & ${{h}_o}$         & Observed wireless channel of one subcarrier \\  \hline
		$n$               & WiFi packet index    &$ \textbf{P}$   &  Multipath profile \\  \hline
		$v_r$             & Path length change rate  &$ {{\textbf{s}_l}} $       &  Steering vector \\  \hline
		$k$               & Subcarrier index     &$ \textbf{H}_{o,0} $  &  Observed wireless channel corresponding to bit 0 of backscatter information \\  \hline
		$f_D$             & Doppler frequency shift     &$ c $  &  Light speed \\  \hline
		$\textbf{D}_r$    & Range estimates from the target to the receiver     &$ \textbf{D}_b $  &  Range estimates from the target to the backscatter tag \\  \hline
		
	\end{tabular}
\end{table}

\begin{figure}
	\centering
	\includegraphics[width=0.5\linewidth]{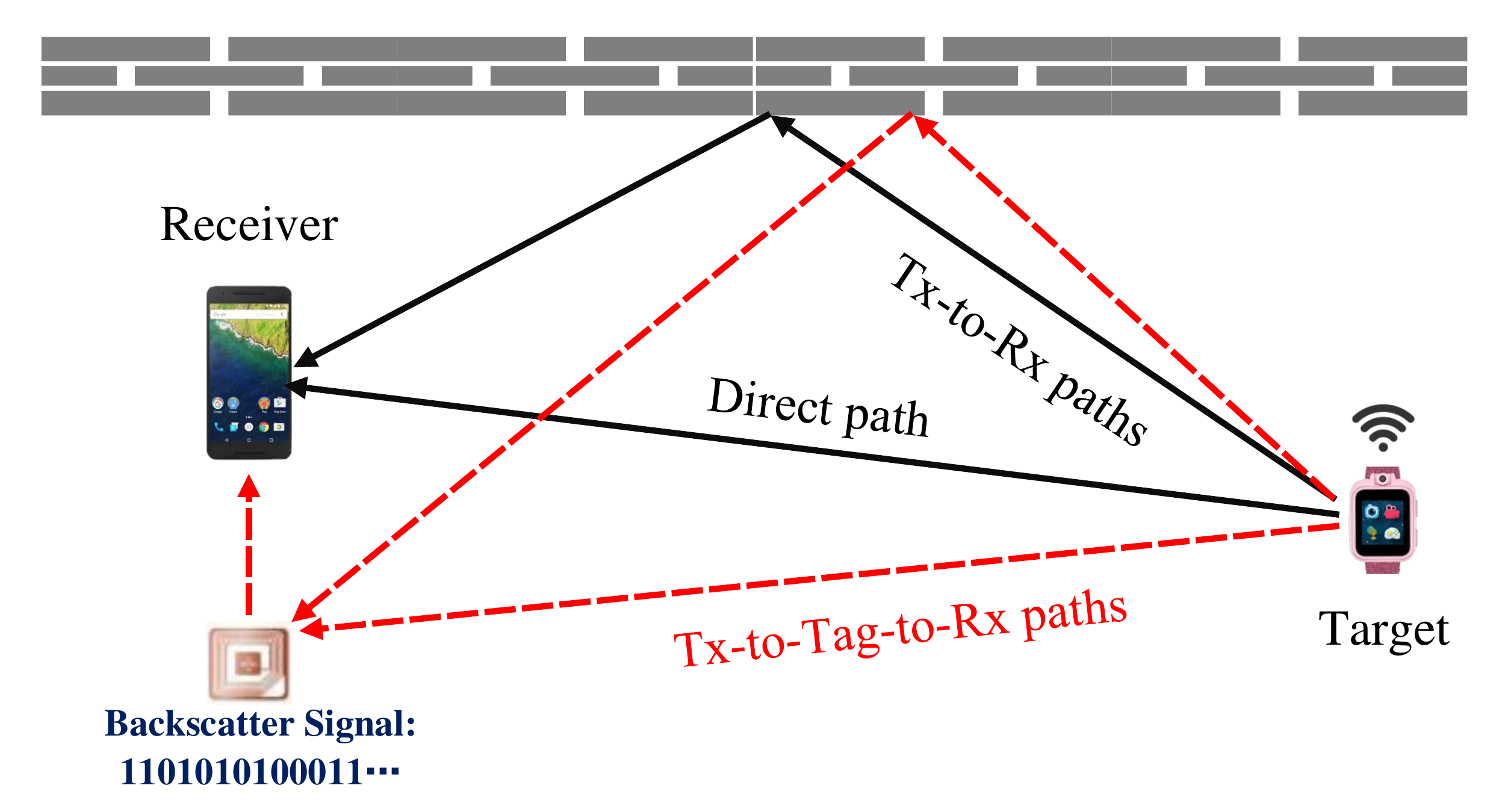}
	\caption{Wireless channel in backscatter networking consists of Tx-to-Rx paths and Tx-to-Tag-to-Rx paths.}
	\label{fig:backscatter-channel-with-tag}
\end{figure}

Specifically, WiFi-based backscatter systems enable a low-power backscatter tag
to convey bits. Smartphones or other wearables with commodity WiFi chips can
receive and decode the backscatter signals. As illustrated in
Fig.~\ref{fig:backscatter-channel-with-tag}, a backscatter tag communicates with
a WiFi device by modulating its wireless channel~\cite{kellogg2014wi}. In
particular, it conveys a sequence of 0 and 1 bits by either reflecting or absorbing the WiFi packets, which changes the wireless channel and causes multipath variation. Mathematically, the wireless channel, ${\textbf{H}_{o}}(n)$, observed on the receiver side for $n$-th packet can be expressed as
\begin{equation} \label{H_tr}
{\textbf{H}_{o}}(n) = {\textbf{H}_{t}}(n) + B(n){\textbf{H}_{b}}(n),
\end{equation}  
where ${\textbf{H}_{t}}(n)$ is the wireless channel between the transceiver pair
without backscatter reflections, $B(n)$ denotes the modulated bits (0 or 1), and ${\textbf{H}_{b}}(n)$ is the wireless channel corresponding to the reflections by the backscatter tag, i.e., the multipath variation.

%Recent progress has explored many ways of resolving the mixed signals. These approaches, however, cannot isolate signals of close-by paths. Note that \sys\ leverages on-body devices for peer-to-peer localization. Thus, the direct path between the transceiver pair and the reflection path of the backscatter sensor are much closer than the resolution limit due to signal bandwidth, e.g., 6~m at 20~MHz. Further the reflected signals are much weaker than the direct-path signal. According to \cite{xiong2015tonetrack}, the smaller peaks corresponding to the reflections will be merged into a larger peak, i.e., the direct-path signal, in the multipath profile, which enforces large error for the parameter estimation of both the direct-path signal and the reflections. 

%To address this, 

\sys's first component focuses on separating the backscatter CSI from the interference due to strong direct-path signal. In principle, one could do that by simply subtracting the CSI measurement corresponding to a bit 0 from that corresponding to a bit 1. However, this solution is complicated by two main factors. First, the backscatter signal is weak, which limits the ability to decode backscatter signal and segment packets corresponding to 0 and 1 bits. Second, due to the lack of tight time synchronization between the transceivers, the observed CSI suffers large time-varying phase distortions, which enforces large errors for the direct subtraction approach. The rest of this section describes how \sys\ addresses these challenges. 

\subsection{Decoding Backscatter Signal} \label{sec:decoding}
To separate the packets corresponding to 0 and 1 bits robustly, \sys\ leverages
the property that each WiFi OFDM symbol has multiple subcarriers. The minimum time interval with which the backscatter tag switches its state is larger than the duration of an OFDM symbol. That is, the duration of each bit transmitted by the backscatter tag is greater than the time it takes to transmit a few OFDM symbols. This ensures that different subcarriers of a single OFDM symbol experience the same backscatter state. Thus, \sys\ can decode the backscatter signal by combing all available subcarriers, which enhances the power of backscatter signals and averages the noise of the CSI measurements. Specifically, for each packet, \sys\ computes the following summation:
\begin{equation}
{\textbf{H}_{c}}(n) = \sum\limits_{k = 1}^K{\left| {h_{o}(k,n)} \right|},
\end{equation}
where ${h_{o}(k,n)}$ is the observed CSI of $k$-th subcarrier, and $K$ is the
total number of subcarriers. Then, similar to \cite{kellogg2014wi}, a moving average method is used in $ {\textbf{H}_{c}}(n) $ where we use a time window with a length of 100~ms to remove the variations in the time domain and normalize the amplitudes. 

\begin{figure} 
	\centering
	\subfigure[Backscatter decoding using a single subcarrier.]{
		\includegraphics[width=0.5\linewidth]{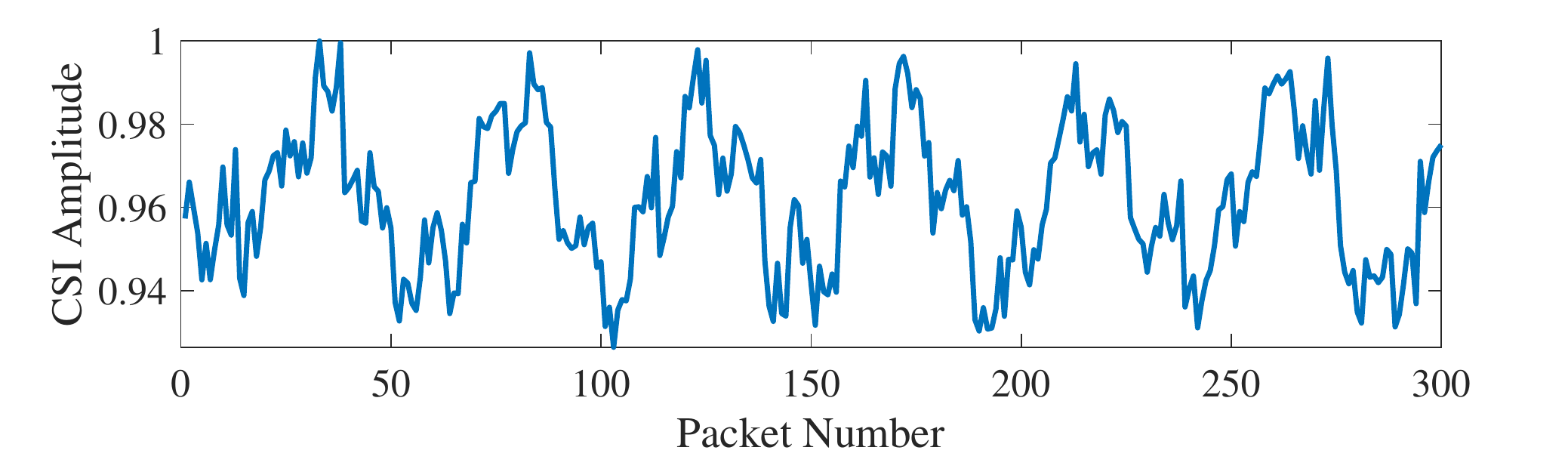}}
	\\
	%\hspace{0.3cm}
	\subfigure[Backscatter decoding using all subcarriers.]{
		\includegraphics[width=0.5\linewidth]{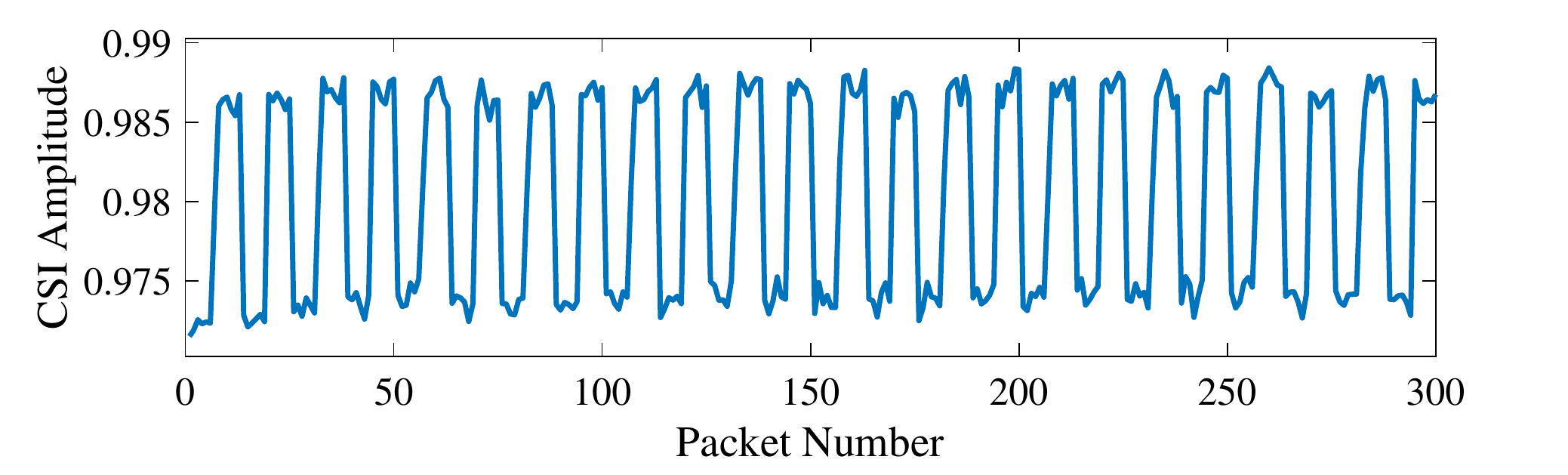}} 
	\caption{Backscatter signal decoding results.}
	\label{SignalDecoding} 
\end{figure}

As shown in Fig.~\ref{SignalDecoding}, the decoded backscatter signal by using
single-subcarrier CSI measurements does not have two distinct levels. In contrast, when combining the CSI measurements of all subcarriers, the backscatter signal can be easily decoded.

\subsection{Handling the Phase Distortions} \label{sec:phase_distortion}
Now that we know how \sys\ decodes backscatter signals, we switch our focus on how to separate the backscatter CSI when diverse phase distortions across packets exist. 

Recall that due to hardware impairment, the observed CSIs are mixed with rich measurement errors from various sources inevitably. For example, sampling frequency offset (SFO) and symbol timing offset (STO) introduce a time delay to the ToF estimation. Carrier frequency offset (CFO) results in a constant frequency shift across all subcarriers, which affects the Doppler estimation. Mathematically, for the $ k $-th subcarrier of the $ n $-th packet, the observed CSI can be expressed as
\begin{equation} 
{{h}_o}(k,n) = {h}(k,n){e^{j(k\varphi+ \theta)}},
\end{equation}
\noindent where $ \varphi $ is the phase distortion caused by STO and SFO, and $
\theta $ is the phase distortion caused by CFO. Next, \sys\ proceeds to
eliminating the effect of CFO, STO, and SFO.

\noindent\textbf{CFO removal.} To remove the effect of CFO, \sys\ leverages the
fact that CFO is frequency-independent~\cite{xie2018precise}. That is, the phase distortion caused by CFO in every frequency band of each packet is identical. Hence, we can use the phase measurement of any subcarrier as a reference to calibrate other subcarriers by simply subtracting from it. In \sys, the phase measured at the first subcarrier is chosen as the reference.

\noindent\textbf{STO and SFO removal.} The phase distortion $  \varphi  $ caused by the STO and SFO is due to the non-synchronized clocks between the WiFi transceivers. Besides, $ \varphi $ is different from packet to packet, since the STO is time-varying. This time-varying phase distortion prevents \sys\ from separating backscatter CSI by directly subtracting the CSI measurement corresponding to a bit 0 from that corresponding to a bit 1.

To eliminate the effect of STO and SFO, conventional approaches rely on CSI
measurements from either multiple channels or multiple antennas. However, it is
inapplicable in peer-to-peer localization where there is only a single antenna
and a single channel measurement. Instead, to retain robust backscatter CSI, we
carefully analyze the effect of STO and SFO in the multipath profile and propose
a novel removal algorithm. The basis of the algorithm is that the multipath
profiles of consecutive packets in coherence time should be identical since the
consecutive packets travel along the same wireless channel. 
%However, as shown in Fig.~\ref{fig:11}, in practice, a time delay exists in the multipath profiles of
%two consecutive packets, which is due to the effect of the time-varying STO and SFO.
It is worth noting that the wireless channel consists of two components, i.e., $ \textbf{H}_{t}$ and $\textbf{H}_{b}$, as described in Section.~\ref{backscatter_communication}. In particular, within the coherence time, the multipath profile of the packet corresponding to a bit 1 has more peaks, i.e., reflected paths of backscatter tag, than that corresponding to a bit 0. Besides, some peaks of the paths from the target to the receiver might merge with the peaks of the reflected paths of the backscatter tag. Thus, there will be errors if we choose the merged peak and simply shift the multipath profile to align this peak. Instead, \sys\ estimates the phase distortion difference by calculating the correlation of two multipath profiles, which leverages the entire multipath profile. Specifically, \sys\ first computes the multipath profile of each packet which will
be introduced in Section~\ref{ToFEstimation}. Let $ \textbf{P}_{n,0}(t) $ and $
\textbf{P}_{n+\Delta n,1}(t) $ be the multipath profile of one packet corresponding to a bit 1 and one corresponding to a bit 0 in coherence time. \sys\ calculates the correlation between these two multipath
profiles by
\begin{equation} 
\textbf{D}(\Delta t) = \sum\limits_{t = 1}^T {{\textbf{P}_{n,0}}(t){\textbf{P}_{n+\Delta n,1}}(t + \Delta t )},
\end{equation}
where $ T $ is the length of $ \textbf{P}_{n,0}(t) $ and $ \textbf{P}_{n+\Delta n,1}(t) $, and $ \Delta t $ is the step length we choose in computing the multipath profile. The difference between STO and SFO, named as $ \tau _{d} $, can be estimated by finding the maximum of $ \textbf{D}(\Delta t) $. Therefore, we can compensate for the effect of STO and SFO by
\begin{equation}
\widehat {{\textbf{H}}}_{o,0}(n) = {\textbf{H}_{o,0}(n)}{e^{j{\pmb{\phi} _{d}}}},\ {\pmb{\phi} _{d}}{\rm{ = }} [2\pi {f_1}{\tau _{d}},2\pi {f_2}{\tau _{d}},...,2\pi {f_K}{\tau _{d}}].
\end{equation}
We repeat this procedure for different pairs of WiFi packets and average the estimates to improve the accuracy. Finally, the backscatter CSI can be separated by subtracting CSI corresponding to a bit 1 from compensated CSI corresponding to a bit 0 as
\begin{equation} \label{backscatterCSI}
\textbf{H}_{b} = \textbf{H}_{o,1} - \widehat{\textbf{H}}_{o,0},
\end{equation}
where $\textbf{H}_{o,1} $ is the observed CSI corresponding to a bit 1, and $\widehat{\textbf{H}}_{o,0}$ represents the CSI corresponding to a bit 0 after STO and SFO compensation.

\begin{figure*}
	\centering
	\begin{minipage}[t]{1\textwidth}\centering
		\subfigure[\scriptsize Multipath profile with phase distortion]
		{\includegraphics[width=0.32\textwidth]{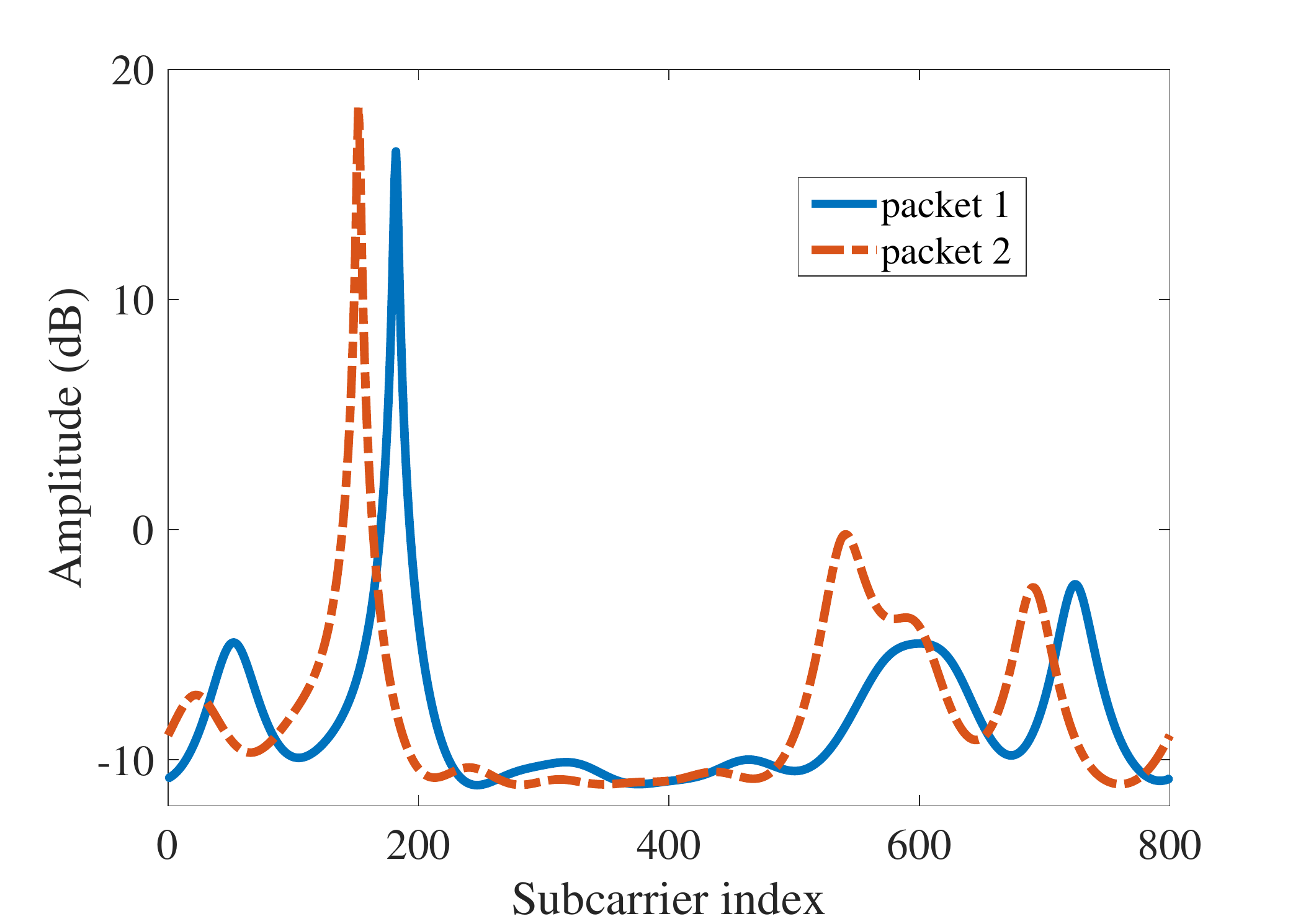}\label{fig:11}}
		\subfigure[\scriptsize Time delay difference estimates]
		{\includegraphics[width=0.32\textwidth]{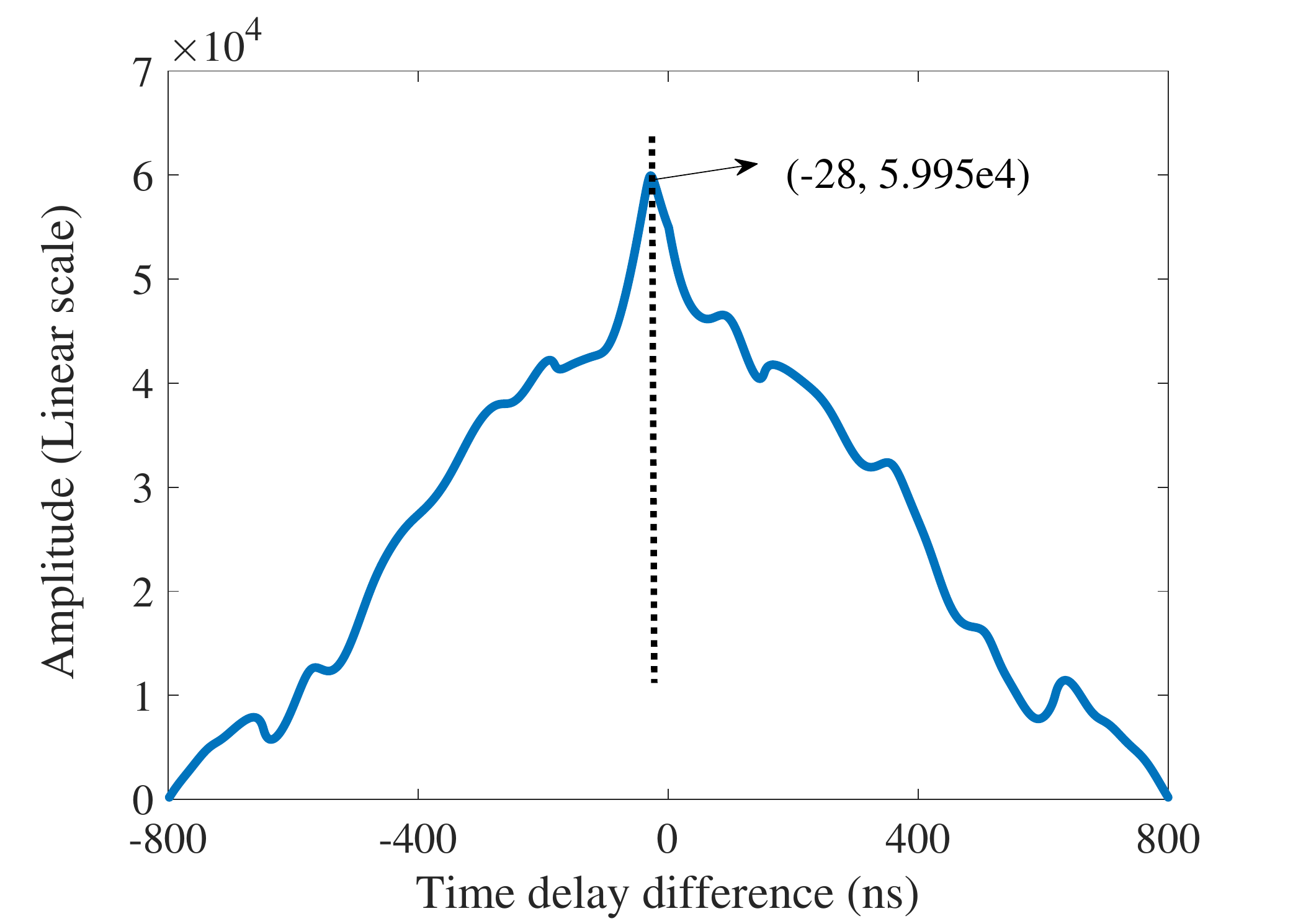}\label{fig:22}}
		\subfigure[\scriptsize Multipath profile without phase distortion]
		{\includegraphics[width=0.32\textwidth]{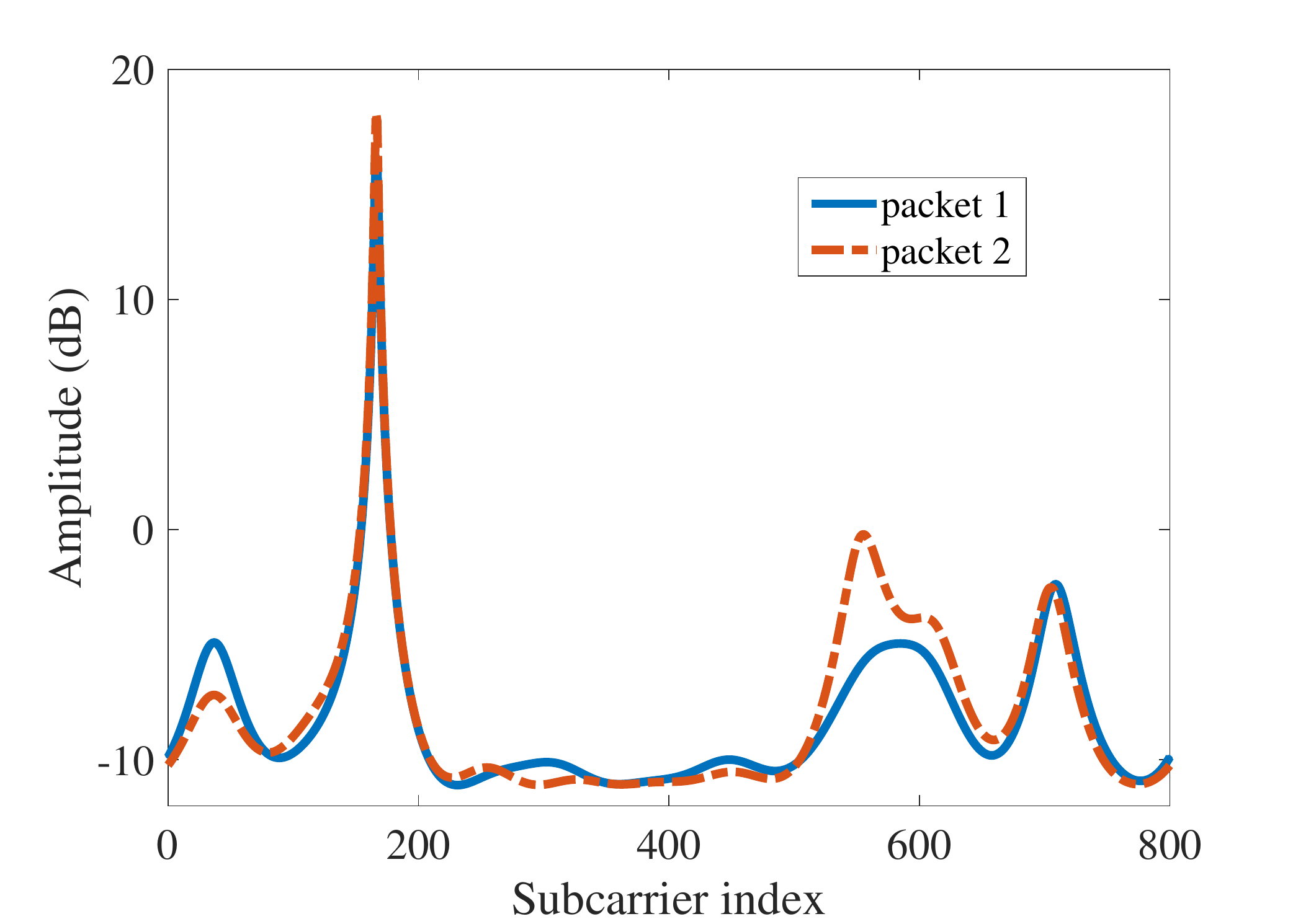}\label{fig:33}}
		\caption{Phase distortions removal. (a) Different phase distortions result in a time delay between two multipath profiles. (b) The time delay can be estimated by comparing these two multipath profiles. (c) After removing the effect of different phase distortions, the corresponding peaks of two multipath profiles coincide with each other.}
		\label{fig:extract}
	\end{minipage} \vspace{0.2cm}
	\hspace{0.2cm}
\end{figure*}

To further illustrate the above algorithm, we collect CSI measurements at 5~GHz
frequency band with the backscatter tag switching between reflecting and
non-reflecting state. We decode the backscatter signal as described in
Section~\ref{sec:decoding} and choose two CSI measurements within
coherence time, one corresponding to a bit 1 of the backscatter information
and the other corresponding to a bit 0. The multipath profiles for the two
CSI measurements are shown in Fig.~\ref{fig:11}. There exists a time
delay between the peaks of these two multipath profiles, which can be determined
by finding the maximum of $ \textbf{D}(\Delta t) $, as depicted in
Fig.~\ref{fig:22}. After removing the effect of STO and SFO, the peaks of two
multipath profiles coincide with each other as shown in Fig.~\ref{fig:33}. Note that after removing the difference of phase distortions between consecutive packets, there exists residual errors. We divide all subcarriers into several groups and leverage the SFO phase error compensation algorithms proposed in~\cite{xie2018precise} to remove residual errors.

\section{Doppler and ToF Estimation} \label{FineDistance}
So far, we have discussed how \sys\ can separate the backscatter CSI. In this section, we describe how it can estimate the Doppler shift and Time of Flight corresponding to the direct path. 

\subsection{Doppler Estimation}\label{sec:doppler}% 1 pp

%\begin{figure}
%	%\begin{figure}[ht]
%	\centering
%	\subfigure[\sys\ leverages user movement to determine the arithmetic sign of Doppler shift.]{\includegraphics[width=0.4\linewidth]{figures/Fig7_1}}
%	\hspace{0.1in}
%	\subfigure[Location errors of virtual antennas have no effect on arithmetic estimation.] {\includegraphics[width=0.4\linewidth]{figures/Fig7_2}}
%	%\hspace{1.7ex}
%	\caption{Illustration of two-step Doppler estimation.}
%	\label{fig:Doppler}
%	%\end{figure}
%\end{figure}

First, we introduce how \sys\ can estimate Doppler shift. Note that \sys\ leverages Doppler shift for fine-grained range and direction estimation. However, it cannot be directly estimated due to the existence of CFO. To address this challenge, \sys\ leverages both amplitude information of CSI and the user movement and proposes a two-step Doppler estimation algorithm.

\noindent\textbf{Doppler effect.} Doppler effect is the frequency change of the signal wave when the transmitter moves relative to the receiver. Let $ f_D $ denote the Doppler shift corresponding to the direct path, $ f $ indicate the center frequency of the signal, and $ c $ represent the speed of light. The radial speed, i.e., the speed along the direct path, can be expressed as
\begin{equation}\label{Eqn:Doppler}
v_r=\frac{f_D}{f}c.
\end{equation}
It is noteworthy that the resolution of Doppler estimation depends on the signal observation interval. For example, at 5~GHz frequency band, a signal observation interval of 1~s corresponds to a resolution of 1~Hz for Doppler estimation, which further corresponds to a resolution of only 0.06~m/s for the path length change rate. Thus, it is fine-grained for range and direction estimation in peer-to-peer localization.

\noindent\textbf{Absolute value estimation.} In practice, due to the existence of CFO which appears as Doppler effect, it is infeasible to directly estimate Doppler shift from the observed CSI measurements. To address this, \sys\ only uses the amplitude information instead of the phase information for Doppler estimation, inspired by prior work~\cite{wang2015understanding,venkatnarayan2019enhancing}. Specifically, we first calculate the CSI power, i.e., $|{{{h}_o}}(k,n)|^2 = |{h}(k,n)|^2$. Then, the short-term Fourier transform (STFT) is applied to compute the spectrogram of Doppler shift corresponding to each subcarrier. It is worth noting that there has no static paths while the receiver or the transmitter moves. Instead, we leverage the observation that there are reflections with very low power which can be regarded as static paths ~\cite{venkatnarayan2019enhancing}. Based on this observation, we can still estimate the Doppler shift while the transmitter and the receiver moves. The signal observation interval is set to 1~s to obtain a resolution of 1~Hz for Doppler estimation as mentioned before. Finally, the radial speed between the receiver and the target can be computed as Eq.~\eqref{Eqn:Doppler}. 

Furthermore, to average the noise and improve the accuracy of speed estimates, we combine the Doppler shift estimated from the CSI measurement of different subcarriers as
\begin{equation}
%v_r= \dfrac{(f_{D_1}/f_1 + f_{D_2}/f_2 + ... + f_{D_K}/f_K)c}{K}.\sum\limits_{k = 1}^K
v_r= \frac{1}{K} \sum\limits_{k = 1}^K \frac{f_{D_k}}{f_k}c.
\end{equation}
Then, we remove the outliers which differ by large variations from the surrounding estimates to obtain the final speed estimates.

\begin{figure}
	\centering
	\includegraphics[width=0.7\linewidth]{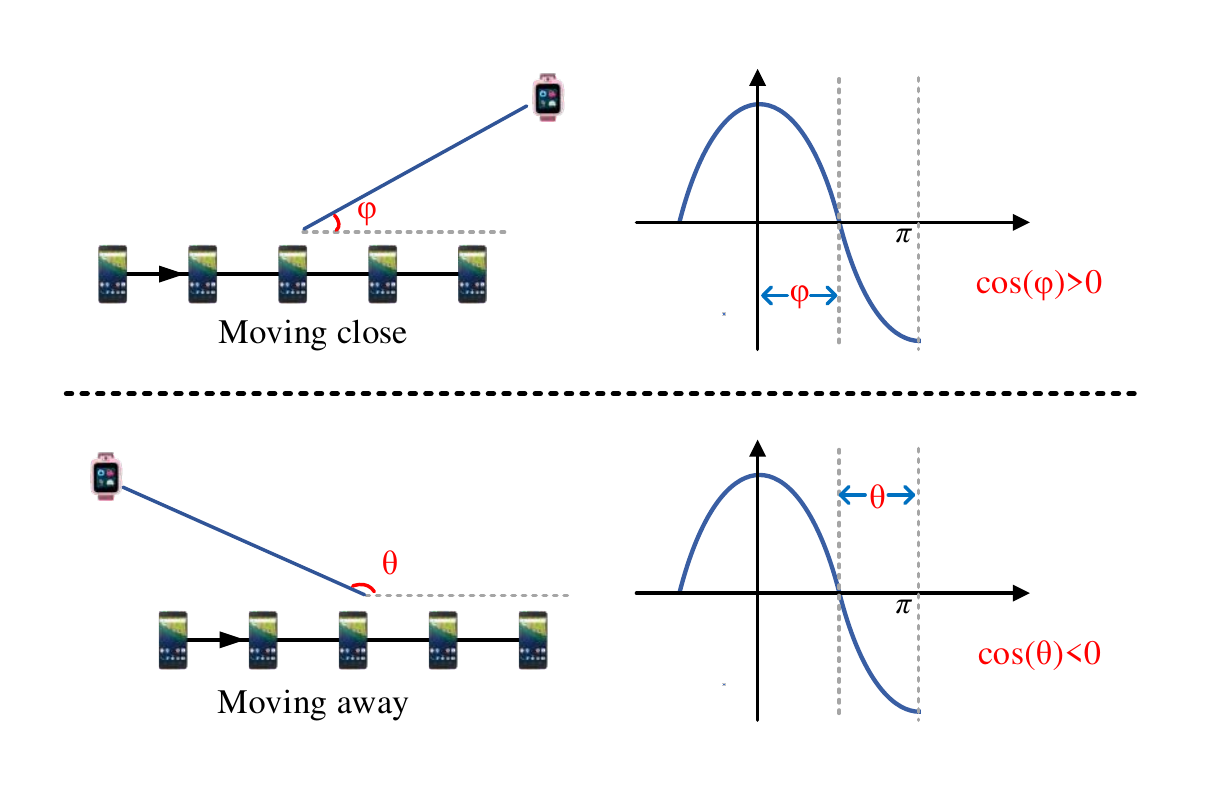}
	\caption{Illustration of Arithmetic sign estimation of Doppler shift.}
	\label{fig:fig71v2}
\end{figure}

\noindent\textbf{Arithmetic sign estimation.} The above approach, however, loses
the arithmetic sign of Doppler shift. That is, with only the absolute value of
Doppler shift, we have no idea whether the user moves away or close to the
target. To address this, \sys\ leverages the synthetic aperture radar (SAR)
created via the user movement. Specifically, according to~\cite{adib2013see}, traditional SAR computes the direction of the signal source by applying standard antenna array equation as % WiVi
\begin{equation}\label{SAR}
{\rm{A[}}\theta {\rm{,}}n{\rm{]  =  }}\sum\limits_{i = 1}^w {h[n + i]{e^{j\frac{{2\pi }}{\lambda }i \Delta \cos \theta }}},
\end{equation}
%\[{\rm{A[}}\theta {\rm{,}}n{\rm{]  =  }}\sum\limits_{i = 1}^w {h[n + i]{e^{j\frac{{2\pi }}{\lambda }i\Delta \sin \theta }}} \]
where $h[n + i] $ is the consecutive channel measurement, $\lambda $ is the
wavelength, and $\Delta $ is the absolute spatial separation between successive
virtual antennas. Note that traditional SAR requires precise locations of virtual antenna
for direction estimation. Unfortunately, such information is unavailable. To address this, we carefully analyze the
effect of errors in location and observe that we can still track the relative
movement, i.e., moving away or close. As illustrated in Fig.~\ref{fig:fig71v2},
when moving close the target, the direction of the target relative to the moving direction of the receiver, i.e., $ \varphi $, varies from $0$ to $\pi/2$, which means $ \cos \varphi > 0$. In contrast, when moving away from the
target, $ \theta $ varies from $\pi/2$ to $\pi$, which means $ \cos
\theta < 0$. Based on this observation, \sys\ determines whether the receiver
is moving close to or away from the target by estimating the sign of the cosine value. Mathematically, let the ground truth spatial
separation be ${\Delta _t}$. Thus, if we use a predetermined value, i.e., $\Delta $, to compute the angle, Eq.~\ref{SAR} can be rewrite as
\begin{equation}\label{SAR_2}
{\rm{A[}}\theta {\rm{,}}n{\rm{]  =  }}\sum\limits_{i = 1}^w {h[n + i]{e^{j\frac{{2\pi }}{\lambda }i \Delta (\frac{{\Delta _t}}{\Delta}  \cos\theta) }}} = \sum\limits_{i = 1}^w {h[n + i]{e^{j\frac{{2\pi }}{\lambda }i \Delta \cos {\theta}_{e} }}}.
\end{equation}
The estimated result is $\cos {\theta _e} = \frac{{{{\Delta _t}}}}{\Delta }\cos \theta $ instead of the truth value $ \cos \theta$. In practice, we cannot obtain the exact value of $\Delta _t $. Fortunately, $ \frac{{{ {\Delta _t}}}}{\Delta } $ is positive, as ${\Delta _t} $ and $\Delta$ represent absolute spatial separation. Thus, $ \frac{{{{\Delta _t}}}}{\Delta } $ does not change the arithmetic sign of final estimates. That is, $ \cos {\theta _e}$ and $ \cos {\theta}$ have the same arithmetic sign, which can still be used to pinpoint the relative movement of the target and the receiver. In \sys, we use the CSI measurements with an interval of 25~ms to synthesize virtual antenna array and set $ \Delta$ as 2.5~cm, as 1~m/s is a reasonable walking speed of human. It is worth noting that, instead of SAR, \sys\ introduces novel algorithms to estimate the precise direction of target.

\begin{figure}
	\centering
	\includegraphics[width=0.5\linewidth]{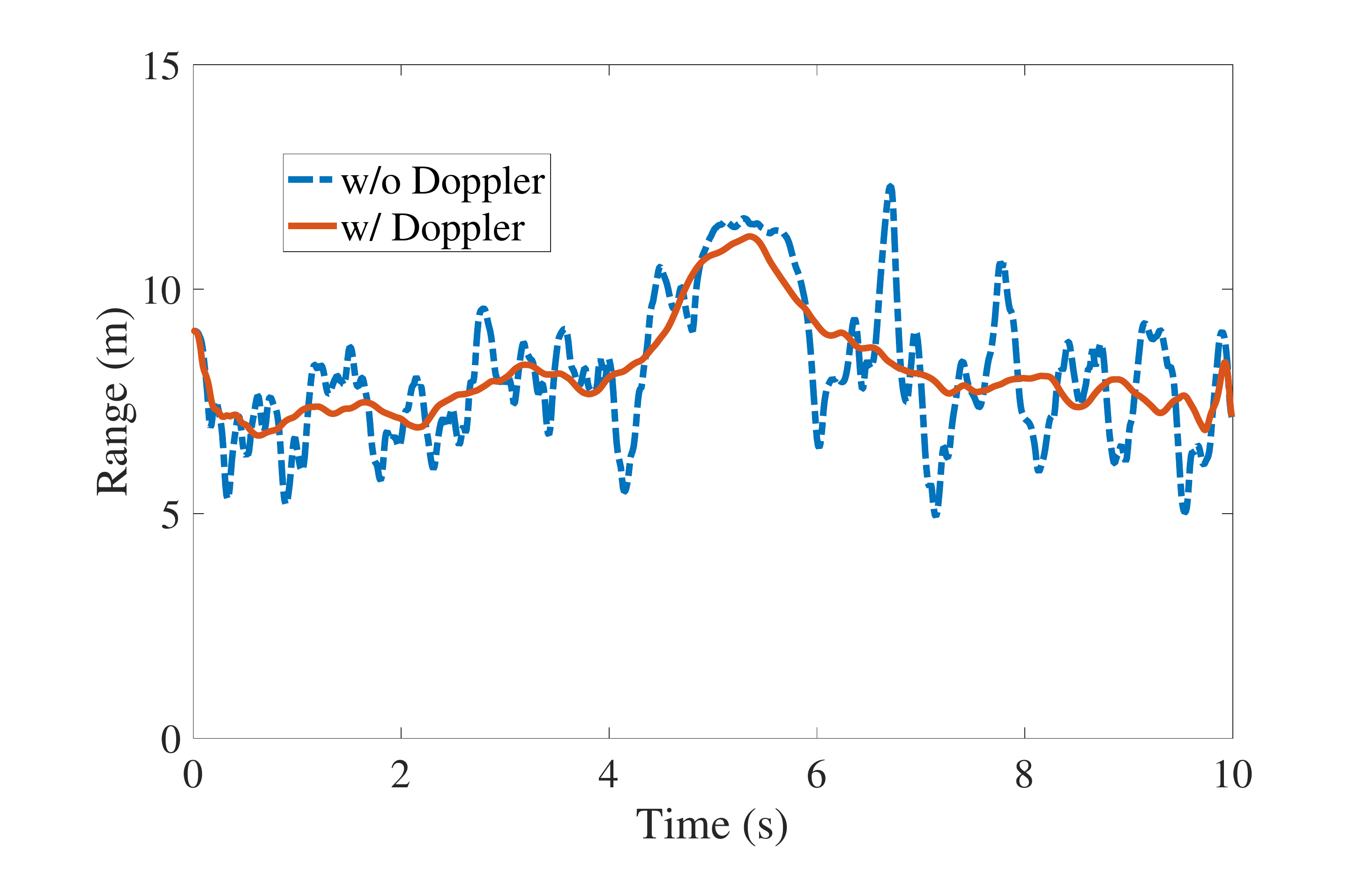}
	\caption{Distance estimation results w/wo fusing Doppler shift.}
	\label{fig:tofrange}
\end{figure}

\subsection{ToF Estimation} \label{ToFEstimation}
Next, we describe how \sys\ estimates the ToF, which is used for range
estimation. Mathematically, as each OFDM symbol has multiple equally spaced
subcarriers, the observed CSI on the receiver can be written in a vector form as
$ \text{\textbf{H}}(n) = [{h}(1,n),{h}(2,n),...,{h}(K,n)] $. To resolve the multipath signal, we leverage the observation that the number of available subcarriers is larger than that of paths in indoor environments. Therefore, we construct a smoothed CSI matrix as a form of Hankel matrix as
\begin{equation}\label{Hankel}
\text{\textbf{H}}(n) = \left[ {\begin{array}{*{20}{c}}
	{{h}(1,n)}&{{h}(2,n)}&{...}&{{h}(K-l+1,n)}\\
	{{h}(2,n)}&{{h}(3,n)}&{...}&{{h}(K-l+2,n)}\\
	\vdots & \vdots & \vdots & \vdots \\
	{{h}(l,n)}&{{h}(l+1,n)}& \ldots &{{h}(K,n)}
	\end{array}} \right],
\end{equation}
where $ l $ is an intermediate parameter and we set $ l=0.5K $ in \sys. According to~\cite{kotaru2015spotfi, gong2018sifi}, the signal space should be orthogonal to the noise space. Thus, we can apply Singular Value Decomposition (SVD) on the smoothed CSI matrix to obtain the noise space $ \textbf{U}_n $ with eigenvalue zero, which is orthogonal to the steering vector $ {{\textbf{s}_l}(\tau)} $, where $ {{\textbf{s}_l}(\tau)} = [\text{e}^{-j2\pi{f_{1}}\tau},\text{e}^{-j2\pi{f_{2}}\tau},...,\text{e}^{-j2\pi{f_{l}}\tau}]^T $. Hence, the ToF estimates can be identified by finding the orthogonal projections of $ \textbf{U}_n $ as
\begin{equation}\label{MUSIC}
\text{P}(\tau ) = \frac{1}{{\left\| {\textbf{U}_n^*{{\textbf{s}_l}(\tau)}} \right\|}}.
\end{equation}
The peaks of the above equation represent the ToF estimates of the direct path and the reflected paths. Typically, there are usually 2 to 5 dominant paths in indoor environments~\cite{kotaru2015spotfi, gong2018sifi, ArrayTrack}. To identify the ToF
corresponding to the direct path between the transceiver pair, we leverage the
intuition that ToF estimates for the direct path will show much smaller
variations compared to the reflections from the environment. That is, the
direct-path ToF estimate from different packets will be clustered together,
while the diameter of each cluster is calculated by the variations in ToF
values~\cite{kotaru2015spotfi}. Specifically, we use the widely adopted Gaussian
Mean clustering algorithm~\cite{kotaru2015spotfi, gong2018sifi, soltanaghaei2018multipath} to eliminate
the multipath effect and identify the direct-path ToF estimates. According to~\cite{kotaru2015spotfi}, the number of cluster is set as 5 and the mean of the tightest cluster is used as the direct-path ToF estimate.
%%spotfi && sifi

It is worth noting that \sys\ only estimates one dimension information, i.e., the ToF. Thus, the computational complexity, which increases exponentially with the number of signal dimensions, is much lower than other localization proposals that estimate parameters from multiple dimensions. Lower computational complexity leads to lower latency for each location estimate in \sys. Furthermore, it is more suitable to deploy \sys\ on on-body devices, such as smartwatches or smartphones, whose computing power is much lower than a dedicated computer.

\section{Peer-to-peer Localization} \label{DirectionEst}
So far, we have discussed how \sys\ can estimate the ToF and Doppler shift. In this section, we describe how it can obtain fine-grained range and direction estimation, and then localize the target.

\subsection{Range Estimation}\label{rangeEst}

To estimate the range, a nature heuristic is to exploit the ToF information
described in Section~\ref{ToFEstimation}. We can compute the range between
transceivers by using the ToF value along with the speed of light. However, the
accuracy of the estimated range suffers from low resolution of ToF due to the
limited signal bandwidth. The blue curve of Fig.~\ref{fig:tofrange} shows the
range calculated directly from the ToF estimates. The range estimates are
fluctuating and thus cannot be directly used for direction estimation and final
localization. 

To refine the range estimation, \sys\ leverages the Doppler shift described in
Section~\ref{sec:doppler}, i.e., the path length change rate. Recall that the
resolution of path length change rate is 0.06~m/s at 5~GHz frequency band with a
signal observation interval of 1~s. It is much fine-grained than ToF estimation,
which has an error of a few ns, leading to meters of ranging error. However,
Doppler shift is relative while the ToF estimation is absolute. To get the best
of both worlds, we incorporate the ToF and the Doppler estimates into a Kalman
Filter to refine the range estimation. The red curve of Fig.~\ref{fig:tofrange}
shows the result after the filter, which is more accurate than the raw
estimation.

\begin{figure}
	%\begin{figure}[ht]
	\centering
	\subfigure[1D case. The moving direction is same as the direction of the line formed by the tag and the receiver.]{\includegraphics[width=0.4\linewidth]{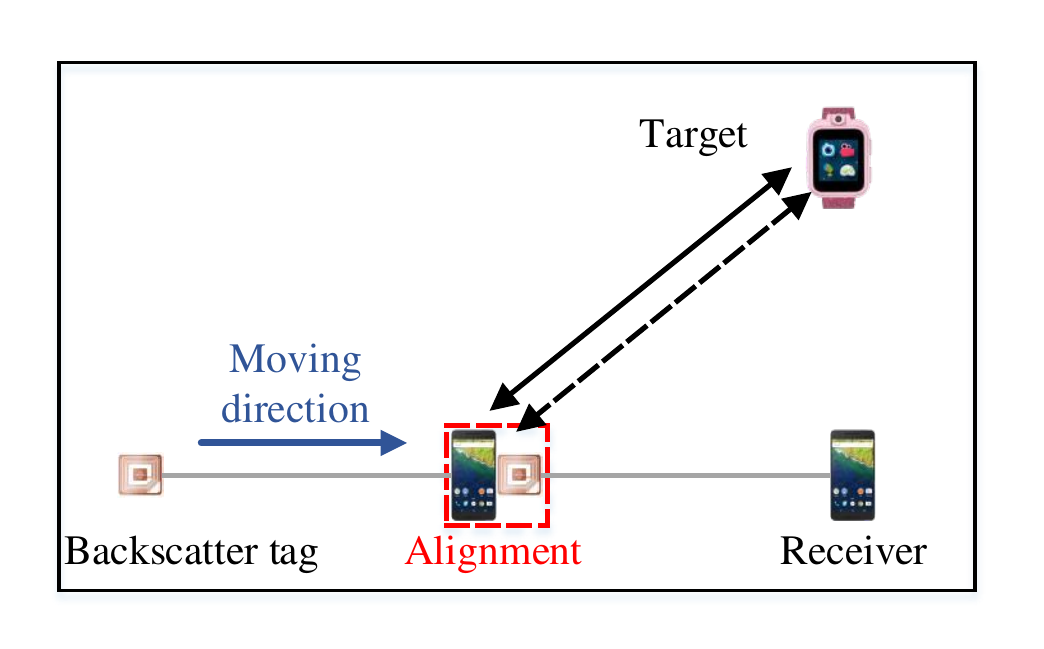}\label{fig:1Dcase}}
	\hspace{0.1in}
	\subfigure[2D case. The moving direction is different from the direction of the line formed by the tag and the receiver.] {\includegraphics[width=0.4\linewidth]{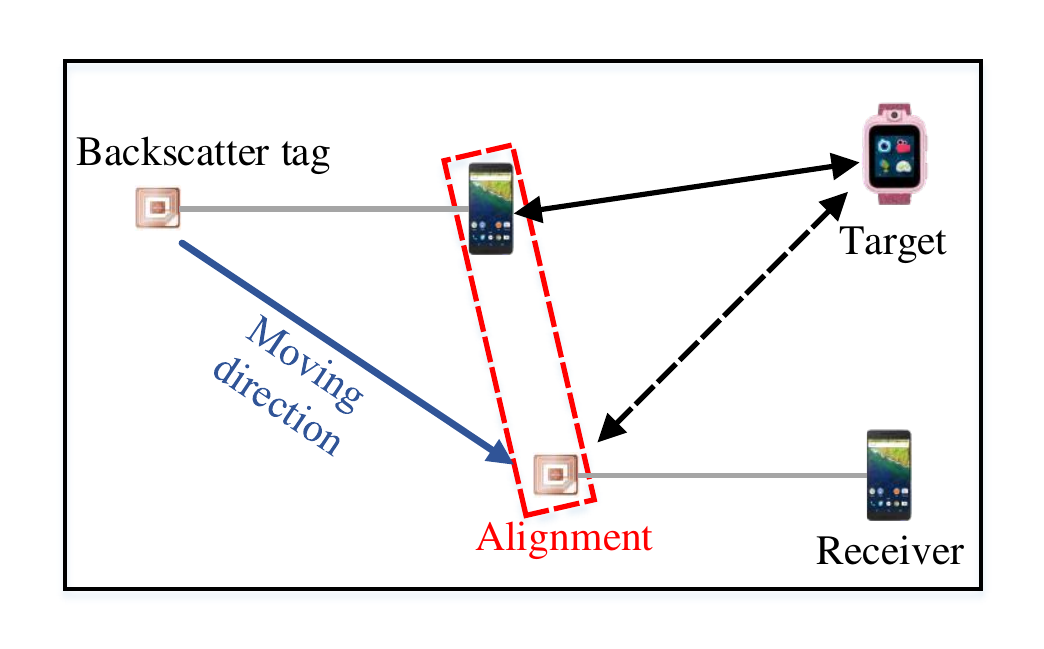}\label{fig:2Dcase}}
	%\hspace{1.7ex}
	\caption{Virtual path alignment.}
	\label{fig:alignment}
	%\end{figure}
\end{figure}

\subsection{Direction Estimation}\label{direction_estimation}
Next, \sys\ proceeds to estimating the direction of the target relative to the line formed by the backscatter tag and the receiver. Our key idea to estimate the direction lies in a novel speed estimation algorithm, named \textit{virtual path alignment}. In this section, we illustrate how \sys\ can estimate the direction in both cases that the target is moving or remains static.

We observe that, in peer-to-peer localization scenarios, the movements of the
target and the receiver are relative to each other. That is, we can consider
that only the receiver is moving, regardless of whether the target is moving or
remains static. As \sys\ leverages the direct-path information that only depends on the relative locations of the target and the receiver for direction estimation, it can also work in the multipath environments. Note that \sys\ extracts the direct-path information in Section.~\ref{ToFEstimation}. Based on this observation, without loss of generality, we
describe the direction estimation algorithm in the case of static target.
Specifically, as the parent carrying the receiver and the backscatter tag moves,
there are two cases as shown in Fig~\ref{fig:alignment}: (i) \textbf{1D case}:
the moving direction is the same as the line between the receiver and the
backscatter tag; (ii) \textbf{2D case}: the moving direction is different from
the line between the receiver and the backscatter tag. Next, we illustrate the
direction estimation, first in 1D and then 2D cases.

\noindent\textbf{1D case.} As shown in Fig.~\ref{fig:1Dcase}, consider a setup
in which the on-body backscatter tag and the receiver, carried by the parent,
are moving along the line formed by themselves towards the target device,
carried by the lost kid. At time $t1$, the backscatter tag arrives at the
location where the receiver traveled through at time $t2$ ($t2<t1$). By
examining the arriving time difference $\Delta t = t1-t2$, the parent's moving
speed can be derived as $v = d_1/\Delta t$, where $d_1$ is the separation
distance between the receiver and the co-located backscatter tag. The value of
$d_1$ can be measured a priori. Therefore, with the radial speed offered by
Doppler estimation, the angle between the moving direction and the line formed
by the target and the receiver can be computed as
\begin{equation}\label{key}
\theta = \arccos{\dfrac{v_r}{v}}.
\end{equation}

Then by continuously aligning the direct-path length corresponding to the
receiver and the backscatter tag, we can obtain the real-time speed of the
parent along the whole trajectory relative to the lost kid, and thus the
real-time direction of the lost kid. However, in 2D case, the above method fails
in estimating the moving speed, because the real moving distance is not $d$. To
address this, we theoretically analyze the effect of the moving direction on the
direction estimation. We verify the moving direction has no effect on the
direction estimation.

%\begin{figure}
%	\centering
%	\includegraphics[width=0.5\linewidth]{figures/2Dcase}
%	\caption{2D case.}
%	\label{fig:2dcase}
%\end{figure}

\begin{figure*}
	\centering
	\begin{minipage}[t]{0.4\textwidth}
		
		\includegraphics[width=1\linewidth]{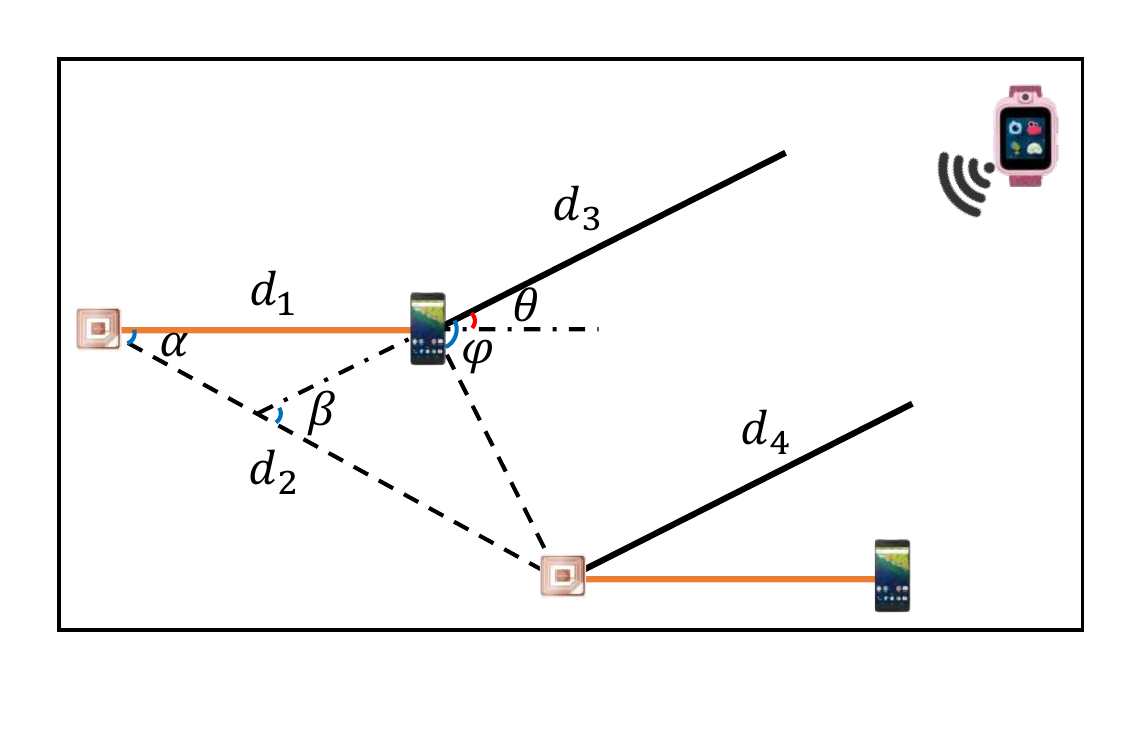}
		\caption{The geometry of 2D case.}
		\label{fig:2DcaseModel}
		
	\end{minipage}\hspace{0.3cm}
	\begin{minipage}[t]{0.4\textwidth}\centering
		
		\includegraphics[width=1\linewidth]{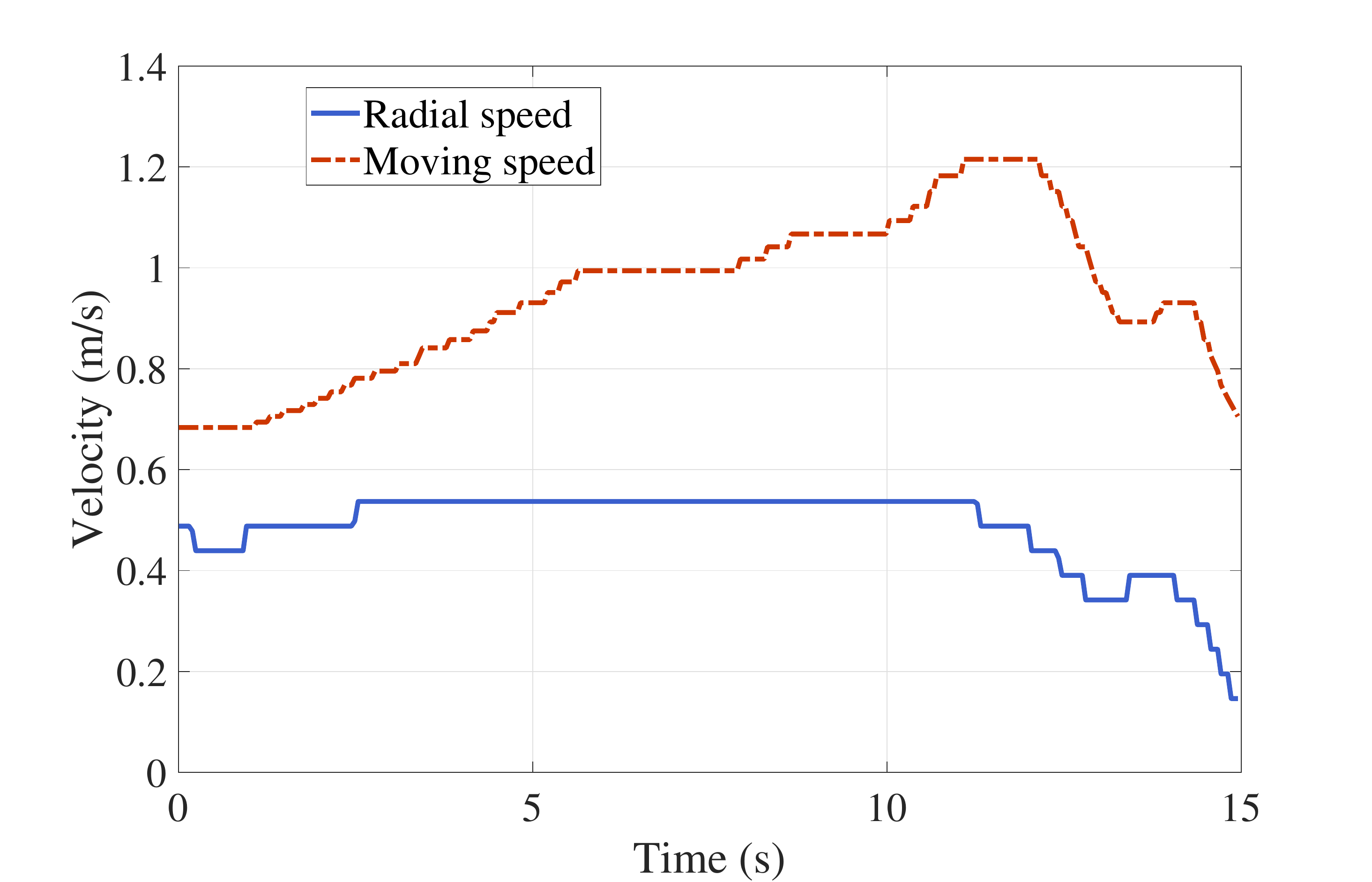}
		\caption{Speed estimation results.}
		\label{fig:speed}
		
	\end{minipage}
\end{figure*}

\noindent\textbf{2D case.} Specifically, without loss of generality, we consider
a setup when the moving direction of the parent is different from the line
formed by the receiver and the backscatter tag. 
As shown in Fig.~\ref{fig:2DcaseModel}, at $ position_1$ at time $ t_1 $, the distance between the receiver and the target is $ d_3 $. when the receiver and the co-located backscatter tag arrive at $ position_2 $ at time $ t_2 $, the distance between the target and the backscatter tag, represented as $d_4 $, equals to the distance between the target and the backscatter tag, i.e., $d_3 = d_4$.
%As shown in Fig.~\ref{fig:2DcaseModel}, when the receiver and the co-located backscatter tag
%arrive at $ position2 $ at time $ t2 $, the distance between the target and the
%receiver at time $ t1 $ equals to the distance between the target and the
%backscatter tag, i.e., $d_3 = d_4$. 
Further, as the distance between the target
and the receiver is much larger than the moving distance within the time
interval of $t_2 - t_1$ and the distance between the receiver and the backscatter
tag, the signals can be considered as parallel incidence. Thus, the direct path
should be perpendicular to the line formed by the position of the receiver at
time $ t1 $ and the position of the backscatter tag at time $ t2 $, i.e., $
\varphi = \pi/2$. In particular, we still use the separation distance between
the receiver and the co-located backscatter tag, i.e., $d_1$, to compute the
moving speed as
\begin{equation}\label{Eq:velocity}
{v_{cal}} = \frac{{{d_1}}}{{\Delta t}} = \frac{{{d_1}}}{{{d_2}}}{v_{real}}.
\end{equation}
where, $d_2$ and $ v_{real} $ are the ground truth moving distance and the moving speed, respectively. Then, according to the triangle theorem, we can obtain the relation of $ d_1 $ and $ d_2 $ as
\begin{equation}\label{Eq:relation}
\frac{{{d_1}}}{{\sin (\frac{\pi }{2} - \alpha  - \theta )}} = \frac{{{d_2}}}{{\sin (\frac{\pi }{2} + \theta )}}.
\end{equation}
Based on Eq.~\ref{Eq:velocity} and Eq.~\ref{Eq:relation}, the cosine value of angle between the transceiver can be estimated as
\begin{align}\label{Eqn:2Dcase}
{\theta _{cal}} &= \arccos{(\frac{{{v_r}}}{{{v_{cal}}}})} \nonumber \\	
&= \arccos{[\frac{{{v_r}}}{{{v_{real}}}}\frac{{cos(\theta )}}{{cos(\alpha  + \theta )}}]}.
\end{align}
Note that $v_r$ corresponds to the radial speed and $ v_{real} $ corresponds to the ground truth moving speed. Thus, ${{{v_r}}} = {{{{v_{real}}}} cos(\alpha  + \theta )}  $. Then, we verified that $\theta _{cal} $ = $\theta $, which indicates that the moving direction have no effect on the direction estimation. Therefore, we can still obtain the direction of target device even use the `wrong' moving distance. To sum up, the proposed direction estimation methods is applicable to the cases where the parent moves, the kid moves, or the parent and the kid move simultaneously. 
%{\color{blue} It is worth noting that the proposed direction estimation methods can work in both obstacle-free and obstacle-rich environments. The reason is that we uses the direct-path information for direction estimation, which only depends on the relative locations of the receiver and the target.}

Next, we describe how \sys\ realizes above idea in practice. Recall that in Section.~\ref{rangeEst}, the fine-grained distance information has been acquired. Let the range estimates corresponding to the receiver and the backscatter tag be $ \textbf{D}_r = [d_{r,t_0}, d_{r,t_1},..., d_{r,t_m}]$ and $ \textbf{D}_b = [d_{b,t_0}, d_{b,t_1},..., d_{b,t_m}] $ respectively. The correlation between $ \textbf{D}_r $ and $ \textbf{D}_b $ can be computed as
\begin{equation} 
{\textbf{C}(\Delta t) = \sum\limits_{t = 0}^T {\textbf{D}_r}(t){\textbf{D}_r}(t + \Delta t )},
\end{equation}
The arriving time difference $ \Delta{t} $ for path alignment between the
receiver and backscatter tag can be estimated by finding the peaks of
$\textbf{C}(\Delta t) $. Finally, the moving speed of the receiver relative to
the target can be estimated as $v = d_1/\Delta t$. Then, we use a moving window
to average the speed estimates, as people move at an even speed during a short
interval. The length of the moving window is set as 600~ms. To further improve
the accuracy of the speed estimates, we remove the estimates that fail to meet
the constraint $ |v|<v_{max} $, as the range of speeds that humans could have in
indoor environments, such as malls or museums, is fairly narrow. 

To illustrate this solution, we measure 5~min CSI data of 40~MHz bandwidth at
5~GHz frequency band, when making the target remain static and the receiver move
in a predetermined trajectory at the speed varying from 0.8~m/s to 1.2~m/s. The distance
between the receiver and the co-located backscatter tag is set as 50~cm. As shown in
Fig.~\ref{fig:speed}, the blue curve is the speed estimate using the above
method and the red curve is the radial speed by Doppler shift. We can observe
that the absolute speed is always larger than the radial speed since the radial
speed is a decomposed component of the moving speed. The estimated moving speed
is varying from 0.7~m/s to 1.3~m/s, which matches the ground-truth speed.

Finally, as the velocity of the moving target is relative constant, we can localize the
target with only on-body single-antenna device by fusing a series of range and direction estimates. 

	\section{Implementation}\label{sec:implementation}%0.5 pp

\begin{figure}
	%\begin{figure}[ht]
	\centering
	\subfigure[Laboratory 1.]
	{\label{Lab1} \includegraphics[width=0.48\linewidth]{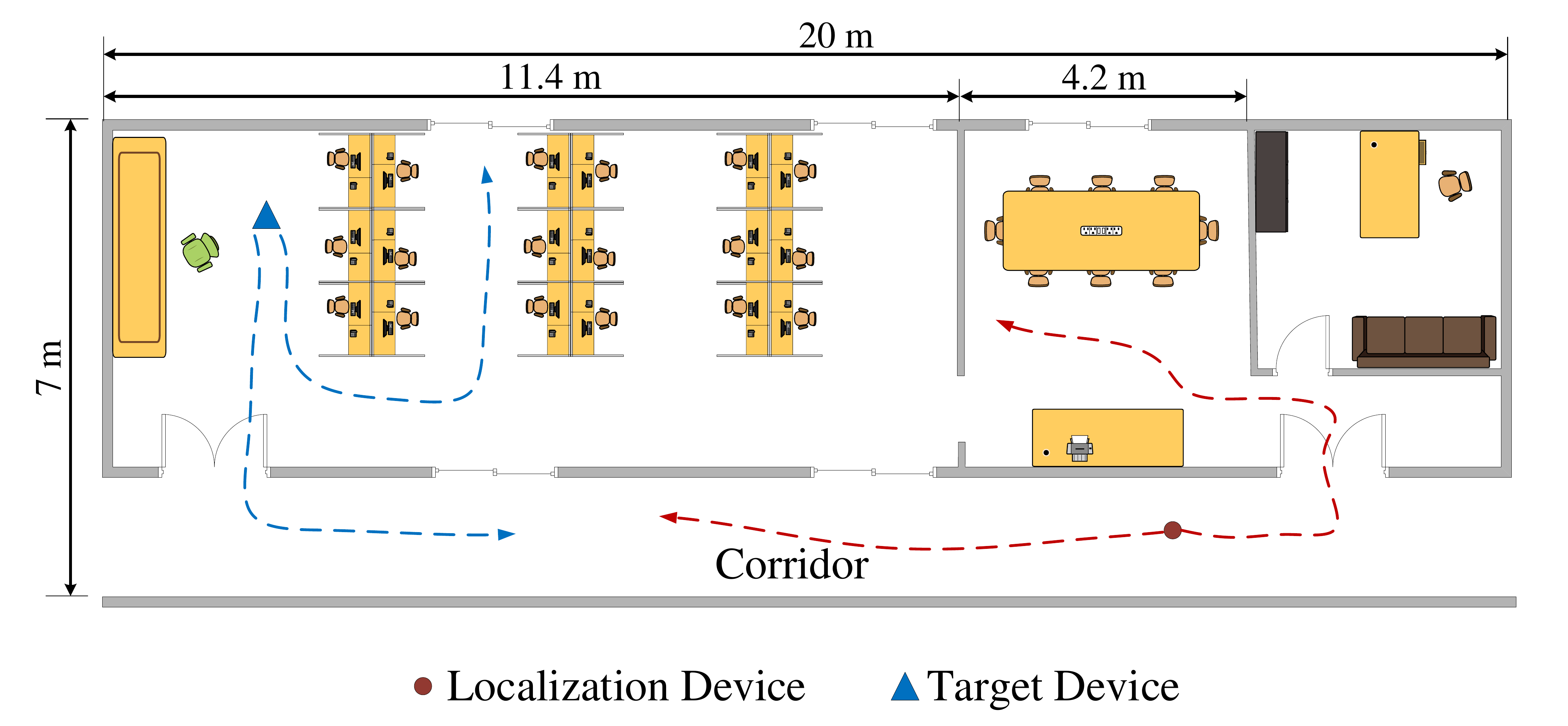}}
	%\hspace{0.1in}
	\subfigure[Laboratory 2.] {\includegraphics[width=0.42\linewidth]{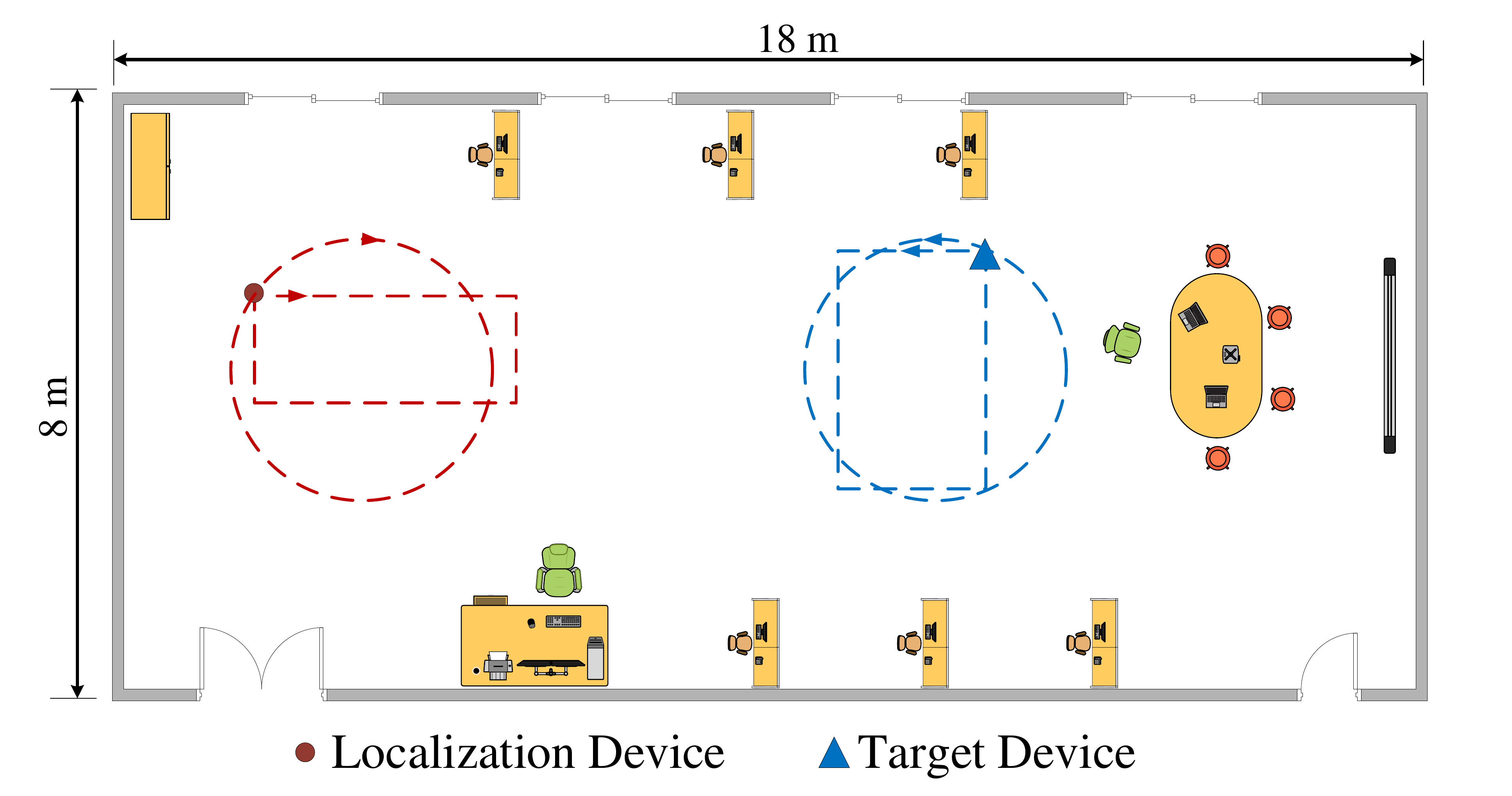}}
	%\hspace{5ex}
	\subfigure[Office building.] {\label{office} \includegraphics[width=0.95\linewidth]{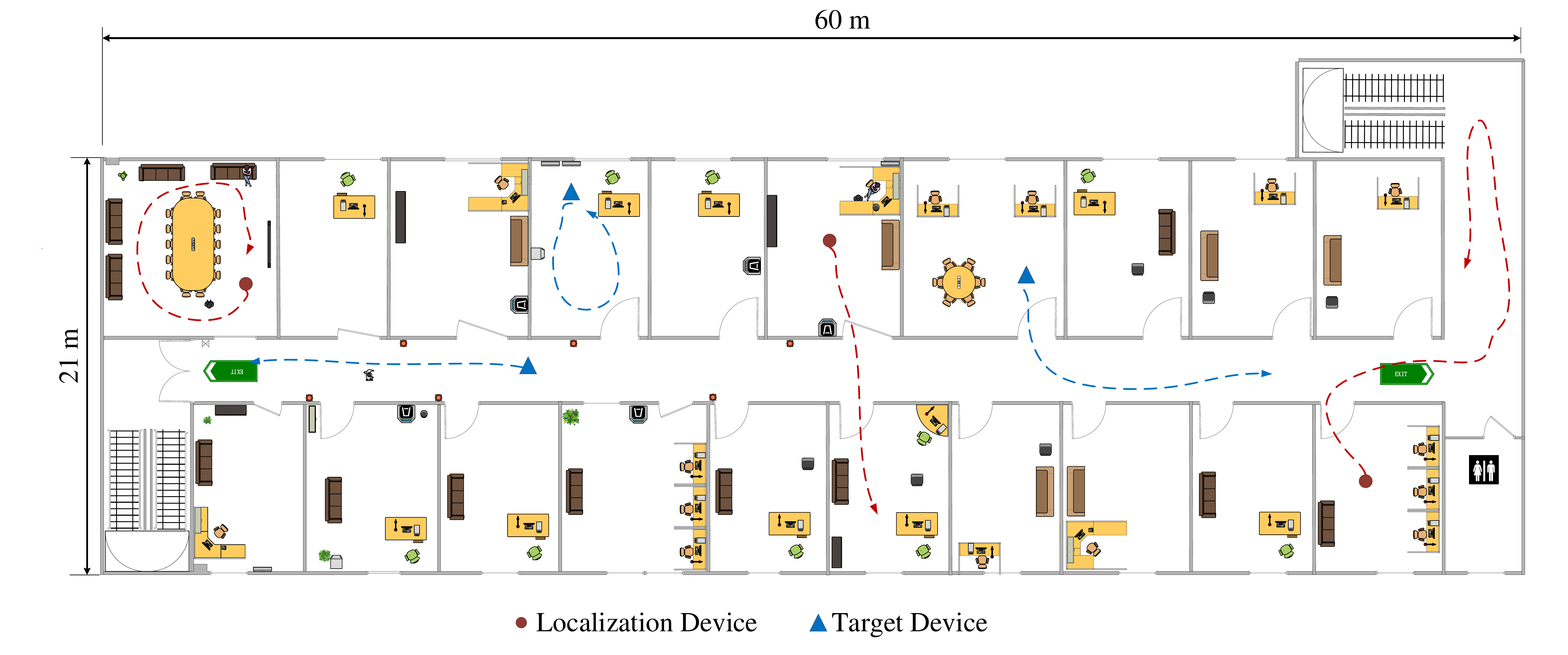}}
	\caption{We test \sys\ over 200 pairs of predefined trajectories in two different settings including two laboratories and an office building. For brevity, part of predefined trajectories are shown in the floor plan where the trajectories of the receiver and the target device are denoted as red and blue dashed lines, respectively. }
	\label{fig:experimental_areas}
	%\end{figure}
\end{figure}

Our \sys\ prototype consists of one customized WiFi backscatter tag and a
pair of WiFi transceivers, which we describe below. 
\subsection{Backscatter Tag}
We implement the backscatter tag based on the open-source FS-Backscatter
hardware~\cite{zhang2016enabling}. The backscatter tag uses an ADG902
transistor to enable backscatter communication. The transistor is connected
to an antenna via an SMA connector and controlled by an Altera STEP-MAX10
FPGA. We program the backscatter tag to transmit data at a bitrate of 1~Kbps
without a frequency shift. The bitrate is carefully chosen based on two
requirements. First, to ensure at least one entire OFDM symbol experiences
the same backscatter state, i.e., the reflecting or non-reflecting state,
the backscatter switching period must be larger than two OFDM symbol
duration. Second, note that the backscatter CSI is separated by using CSI
measurements corresponding to 0 and 1 bits. To ensure the wireless
channel remains stable while separating backscatter CSI, the backscatter
switching period must be smaller than half the duration of coherence time. To this end, we program the backscatter to transmit data at a bitrate of 300~bps, while the WiFi packet rate of the transmitter is set to 1~kHz.

\subsection{Transceiver}
%Note that \sys\ can enable many applications, 
%such as finding a lost kid, tracking goods and relocating medical equipment. Therefore, 
Considering the diverse use cases of \sys, 
we implement the transceiver using three different types of off-the-shelf commodity devices, including Intel 5300 WiFi card representing commonly used APs, Raspberry Pi B4 representing programmable devices and Google Nexus 6p representing on-body devices. All three devices can provide CSI measurement for each WiFi packet.

\noindent\textbf{Intel 5300 WiFi card.} We mount Intel 5300 WiFi cards on two Intel NUCs D54350WYKH with a 1.3~GHz Core i5 processor with 4 cores, a 120~GB SSD and an 8~GB RAM, running the Ubuntu 16.0.4 operating system. They are connected to an omni-directional VERT2450 antenna and serve as the transmitter and the receiver, respectively. We use the Linux 802.11n CSI Tool~\cite{halperin2011tool} to obtain the CSI measurements of each WiFi packet from both the transmitter and the backscatter tag. Both the transceiver operate on 5~GHz WiFi spectrum due to the firmware limitations mentioned in ~\cite{phaser}.

\noindent\textbf{Raspberry Pi B4 and Google Nexus 6p.} For real-world deployment, we also employ Google Nexus 6p and Raspberry Pi B4 for evaluation, which are equipped with bcm4358 and bcm43455c0 WiFi chip, respectively. We use the Nexmon CSI Extractor Tool \cite{schulz2018shadow} for CSI collection, which allows per-frame CSI extraction for up to 80~MHz bandwidth in both 2.4~GHz and 5~GHz frequency bands. It supports up to nine Broadcom WiFi chips and allows \sys\ to extract CSI at 64 subcarriers. Each of the real and imaginary parts of CSI for each subcarrier is represented by using 9 or 12 bits for different types of WiFi chips.

\subsection{Software}
For Intel 5300 WiFi cards and Raspberry Pis, \sys's localization algorithms
are executed locally. For Google Nexus 6p, \sys's localization algorithms
are executed on a workstation with an Intel Core i7-6700K 4.2~GHz CPU and 16~GB RAM. The CSI measurements and IMU sensor data, used for IMU-assisted localization method, are transferred via LTE connections. All algorithms of \sys\ are implemented using MATLAB in our current prototype.	

%We adopts two USRP X310 nodes with CBX daughterboards as the transmitter and receiver and Both USRPs are implemented with the WiFi 802.11n protocols. each USRP is connected to an omni-directional VERT2450 antennas. The USRPs sample at 20/40 MSps and send data over Ethernet to a computer using the Intel Converged Network Adapter X710-DA2 to support high data rates. The received signals are processed in Matlab using a comupter which has an 4-core 8- thread 64-bit Intel Core i7 processor and 16GB RAM.

%\textbf{Backscatter tags} The WiFi-based backscatter tags are customized according to ~\cite{zhang2016hitchhike}. We use three to six backscatter tags as reference tags to help lcoalize the target.
%\begin{figure}
%	\centering
%	\includegraphics[width=0.33\linewidth]{figures/setup}
%	\caption{On-body device placement.}
%	\label{fig:setup}
%\end{figure}

\begin{figure*}[t]
	\centering
	\begin{minipage}[t]{0.35\textwidth}\centering
		\includegraphics[width=1\textwidth]{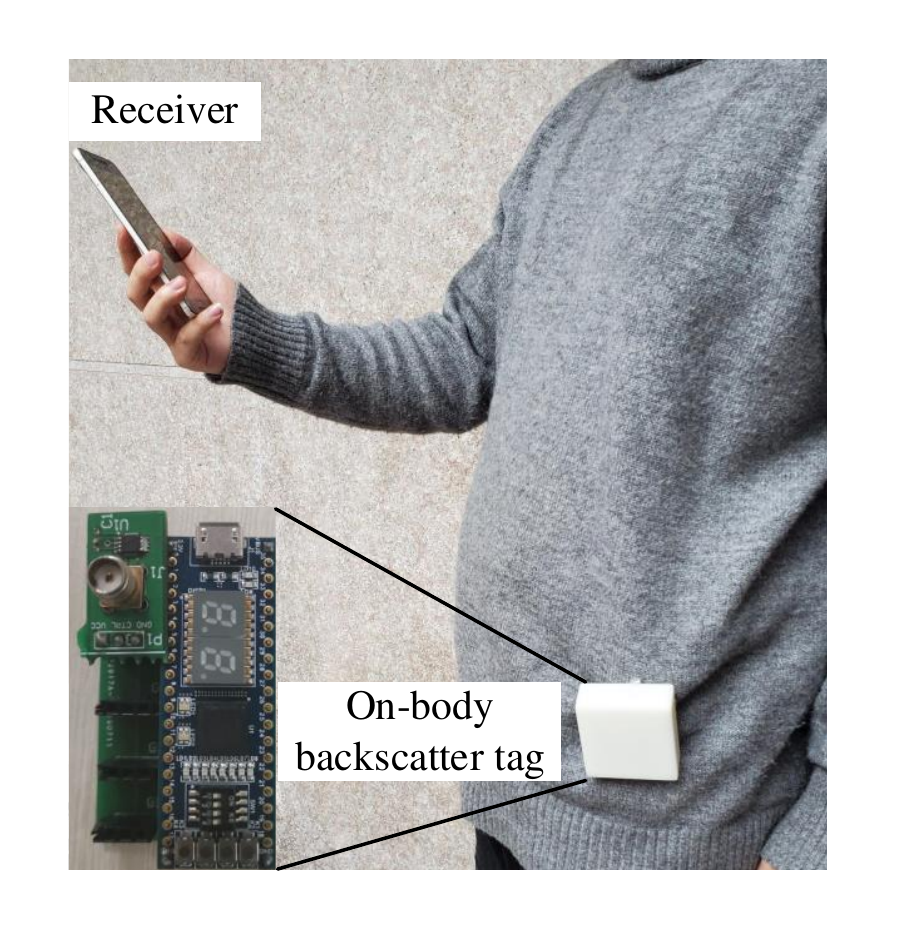}
		\caption{On-body backscatter tag placement.}
		\label{fig:setup}
	\end{minipage} \hspace{0.2cm}
	\begin{minipage}[t]{0.55\textwidth}\centering
		\includegraphics[width=1\textwidth]{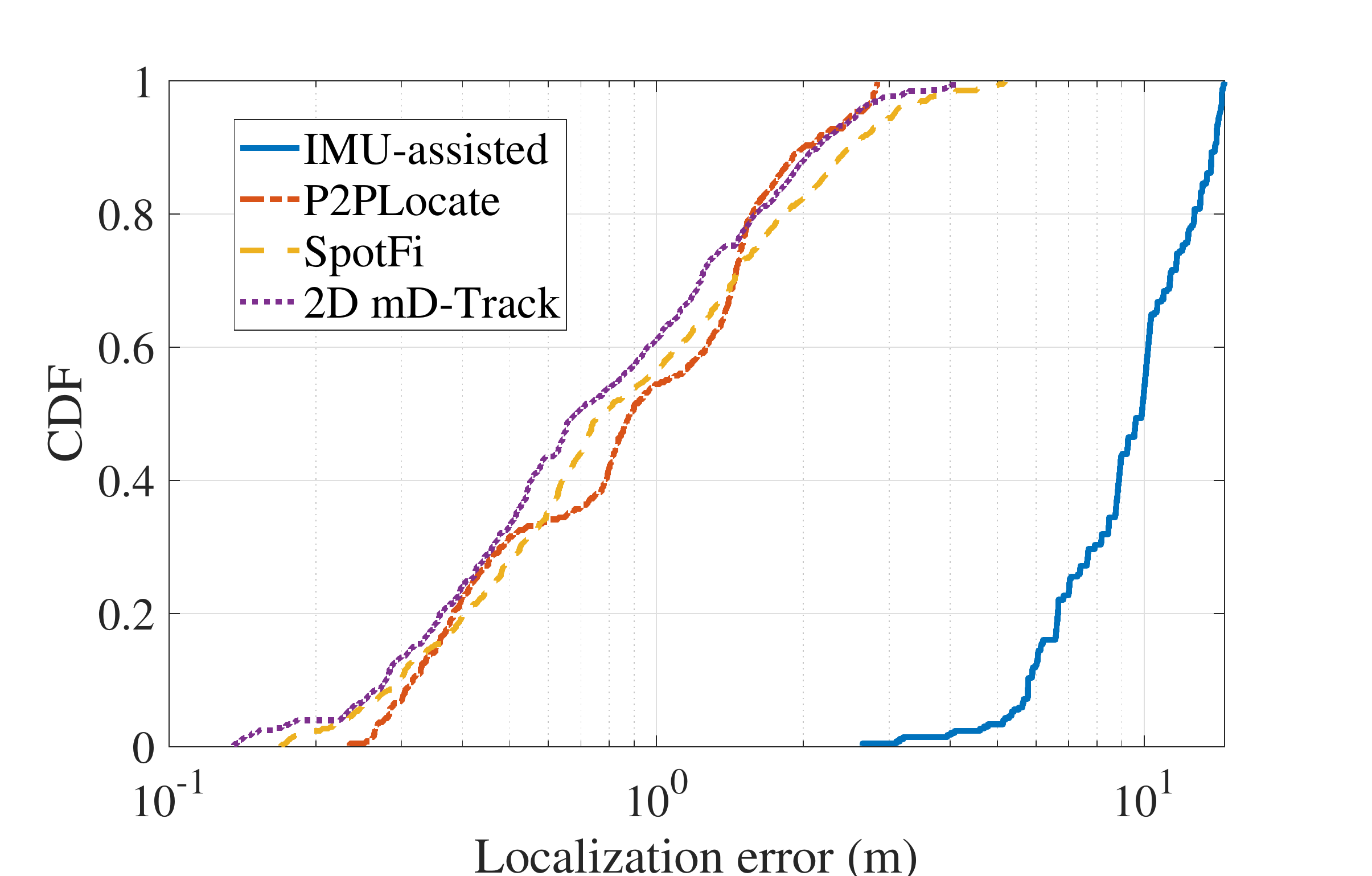}
		\caption{Localization error with static target.}
		\label{fig:distanceerror}
	\end{minipage}  
\end{figure*}

%    \begin{figure}
%    	\centering
%    	\includegraphics[width=0.45\linewidth]{figures/DopplerDirection}
%    	\caption{Arithmetic sign estimation accuracy of Doppler shift.}
%    	\label{fig:dopplerdirection}
%    \end{figure}

%    \begin{figure}
%    	\centering
%    	\includegraphics[width=0.97\linewidth]{figures/floorPlan0720}
%    	\caption{Experimental floorplan.}
%    	\label{fig:floorplan0720}
%    \end{figure}

%    \begin{figure}
%    	\centering
%    	\includegraphics[width=0.97\linewidth]{figures/yinjian}
%    	\caption{Hardware}
%    	\label{fig:yinjian}
%    \end{figure}

\section{Evaluation}\label{sec:evaluation}%0.5 pp
In this section, we conduct extensive experiments in various environments to evaluate the performance of \sys. We first present the experimental setup and then introduce the evaluation results.	

\subsection{Experimental Setup}
We conduct extensive experiments in two different settings: (i) two
different floors of a laboratory measuring 7~m by 20~m and 8~m by 18~m,
respectively; (ii) an office area measuring 21~m by 60~m. As shown in
Fig.~\ref{fig:experimental_areas}, these areas have diverse multipath
characteristics and different floor layouts. Besides, during our experiments, there are around 25 and 15 individuals in the laboratory 1 and laboratory 2, respectively. In the office area, there is an average of 3 to 5 individuals in each room and tens of individuals that are walking around. In particular, we deploy four additional receivers equipped with three antennas each to demonstrate the performance of infrastructure-based localization system. 

Our experiments are conducted without changing user behaviors. As showed in Fig.~\ref{fig:setup}, one volunteer, serving as the localization end, holds a receiver and a customized backscatter tag is attached to his waist. Another volunteer, serving as the target, puts the transmitter in his pocket. During the experiments, the localization end walks along predefined trajectories with diverse speed. The target either stands still or walks along predefined trajectories to evaluate \sys's performance for both cases of static and moving target. Note that we test over 200 pairs of predefined trajectories of the receiver and the target in each settings. For brevity, we show only a part of the predefined trajectories in Fig.~\ref{fig:experimental_areas}.

\noindent\textbf{Ground truth.} We obtain the ground truth via a
camera-based tracking system along with the floor plan. Specifically, we
first define several trajectories for evaluations. Then, several digital
cameras are installed along the predefined trajectories and capture user
movement in real time. Finally, to match the location estimates of the camera-based tracking system with the corresponding CSI measurement at that location, we hand-labeled all video date for ground truth by comparing the timestamps with the RF samples.

\noindent\textbf{Baseline.} We compare \sys\ with three localization systems, i.e., IMU-assisted system~\cite{venkatnarayan2019enhancing}, SpotFi~\cite{kotaru2015spotfi} and mdTrack~\cite{xie2019md}. We choose the IMU-assisted system as it best fits our experimental setup, where both the target and the receiver only have a single antenna. Note that other peer-to-peer localization schemes like Chronos~\cite{vasisht2016decimeter} and MonoLoco~\cite{soltanaghaei2018multipath} require multiple antennas in both the receiver and the target. Additionally, ppNav~\cite{yin2017peer} enables peer-to-peer navigation using smartphones, it relies on the wireless fingerprints offered by multiple infrastructures. These systems fail to work properly in our peer-to-peer scenario, and thus we cannot fairly compare our system with them.

%because they require multiple antennas which is not applicable in our scenario. Note that WiTag~\cite{kotaru2017localizing} mentioned in the comment as [1] leverages multiple three-antenna APs to localize low-power backscatter tags which is not applicable in peer-to-peer localization scenario. ppNav~\cite{yin2017peer} mentioned in the comment as [2] enables peer-to-peer navigation using smartphones by utilizing wireless fingerprints, which requires multiple APs.

%We choose IMU-assisted localization system because it can directly operate on single-antenna mobile devices without hardware modification. We don't compare \sys\ with other peer-to-peer localization system, like Chronus and MonoLoco, as they require multiple antennas in both the receiver and the target. While ppNav enables peer-to-peer navigation using smartphones, it relies on both wireless and vision information. Hence, we don't use them for comparison.}

We deploy an IMU-assisted localization system as baseline, which integrates wireless signals and IMU
data to localize the target. Specifically, since the mobile devices are
equipped with IMU sensor, we collect both the CSI measurements and IMU data
with time stamps. Then, the IMU data is used to estimate the moving speed of
the receiver and keep the rest unchanged compared to \sys. We also compare
\sys\ with two state-of-the-art infrastructure-based localization systems, i.e., SpotFi~\cite{kotaru2015spotfi} and mD-Track~\cite{xie2019md}. We deploy four receivers with three antennas each in the environment. Then, we compute angle of arrival (AoA) of the target by using the algorithm proposed in SpotFi and mD-Track, respectively. Finally, we localize the target by triangulating the AoA measurements across the receivers.

\begin{figure*}[t]
	\centering 
	\begin{minipage}[t]{0.45\textwidth}\centering
		\includegraphics[width=1\textwidth]{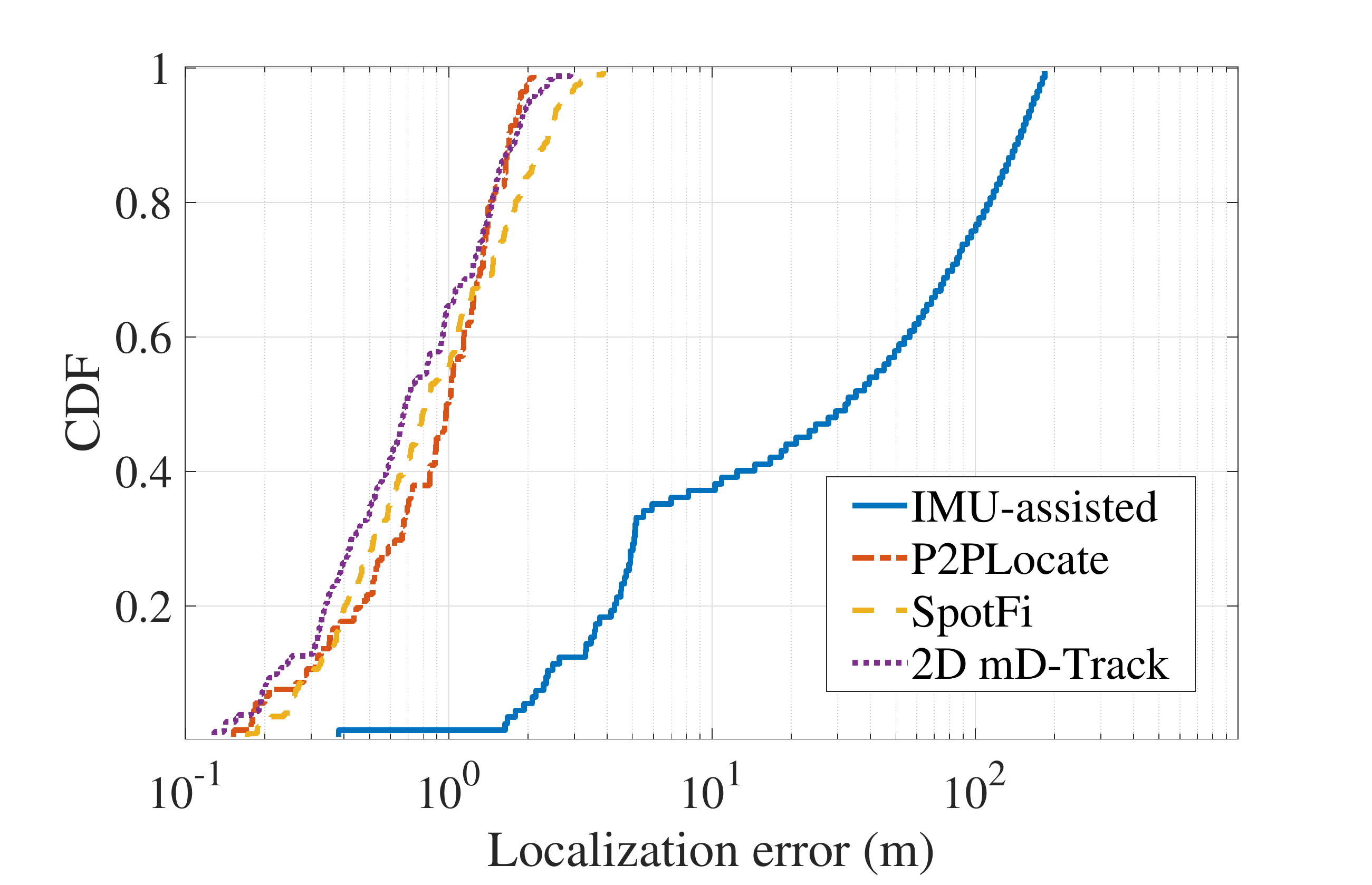}
		\caption{Localization error with moving target.}
		\label{fig:distanceerror_motion}
	\end{minipage} \hspace{0.2cm}
	\begin{minipage}[t]{0.45\textwidth}\centering
		\includegraphics[width=1\textwidth]{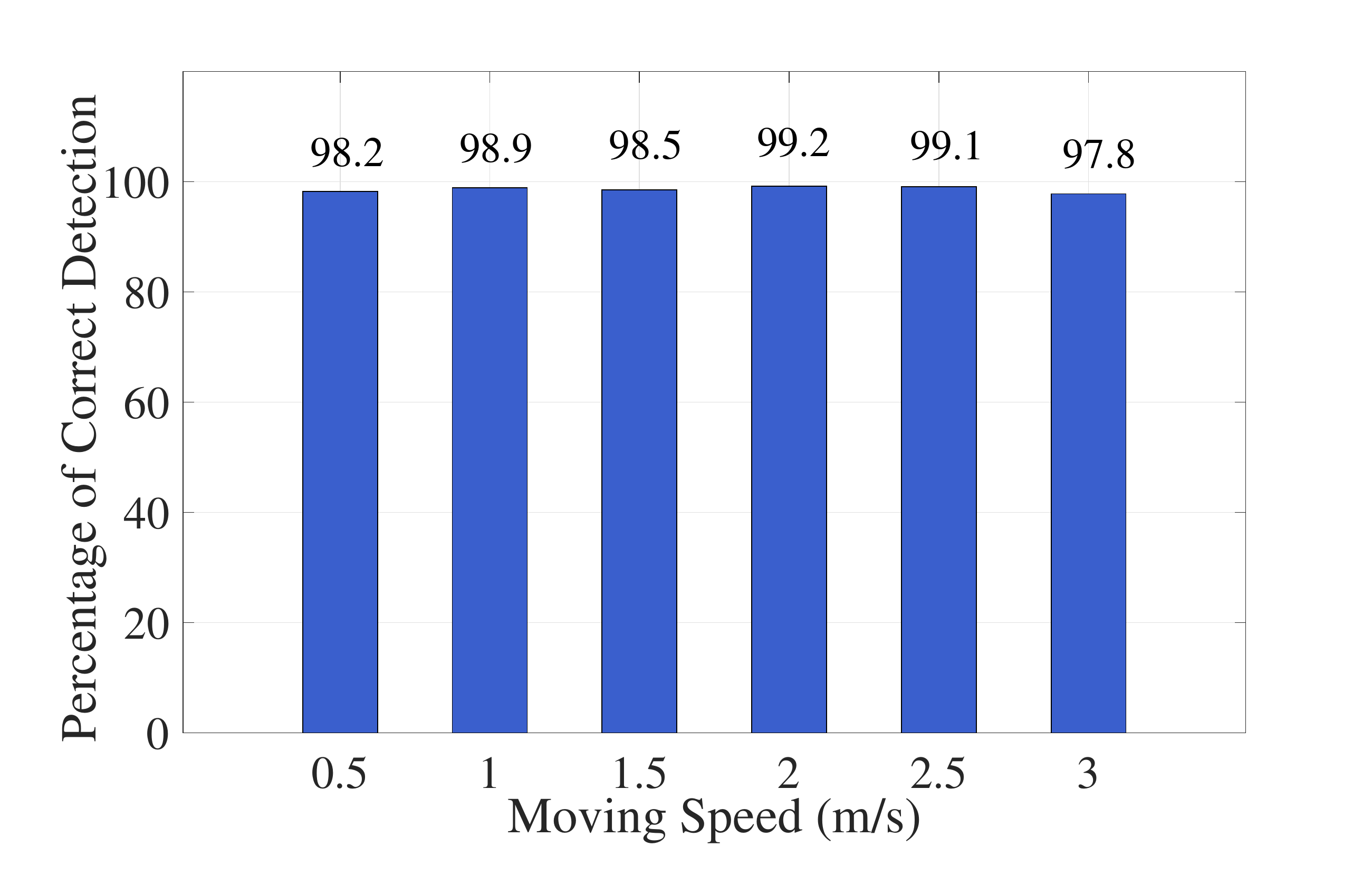}
		\caption{Arithmetic sign estimation accuracy of Doppler shift.}
		\label{fig:dopplerdirection}
	\end{minipage} 
\end{figure*}

\subsection{Overall Accuracy} \label{overallAcc}
We start by presenting the overall accuracy of \sys\ and comparing it with IMU-assisted and infrastructure-based localization methods. We evaluate \sys\ with both static and moving targets. To calculate localization error, we measure the Euclidean Distance between the estimated locations and the ground truth. Then, we use the Cumulative Distribution Function (CDF) of the localization error to demonstrate the performance.

\noindent\textbf{Static target.} Fig.~\ref{fig:distanceerror} shows the
localization results in the case where only the localization end moves.
\sys\ achieves a median localization accuracy of 0.88~m compared to 9.9~m
for IMU-assisted solution, 0.79~m for SpotFi and 0.68~m for 2D mD-Track. The 90th percentile tail
errors are 1.98~m, 13.97~m, 2.43~m and 2.11~m for \sys, IMU-assisted solution,
SpotFi, and 2D mD-Track, respectively. Thus, \sys\ achieves similar accuracy and reliability
compared to state-of-the-art infrastructure-based localization systems using only single-antenna on-body devices, making it possible to enable emerging applications mentioned in Section~\ref{sec:Introduction}.

The primary reason for the large errors for IMU-assisted solution is that the IMU data has large accumulative errors since it requires integration to compute speed. SpotFi and mD-Track achieve more accurate localization results because it leverages spatial information offered by not only multiple receivers but also multiple antennas. However, the accuracy does not improve a lot because it requires the existence of LoS path between the target and all receivers, which limits its usage. To the best of our knowledge, no other localization system achieves decimeter-level accuracy when using only a single antenna.

\begin{figure*}[t]
	\centering
	\begin{minipage}[t]{0.45\textwidth}\centering
		\includegraphics[width=1\textwidth]{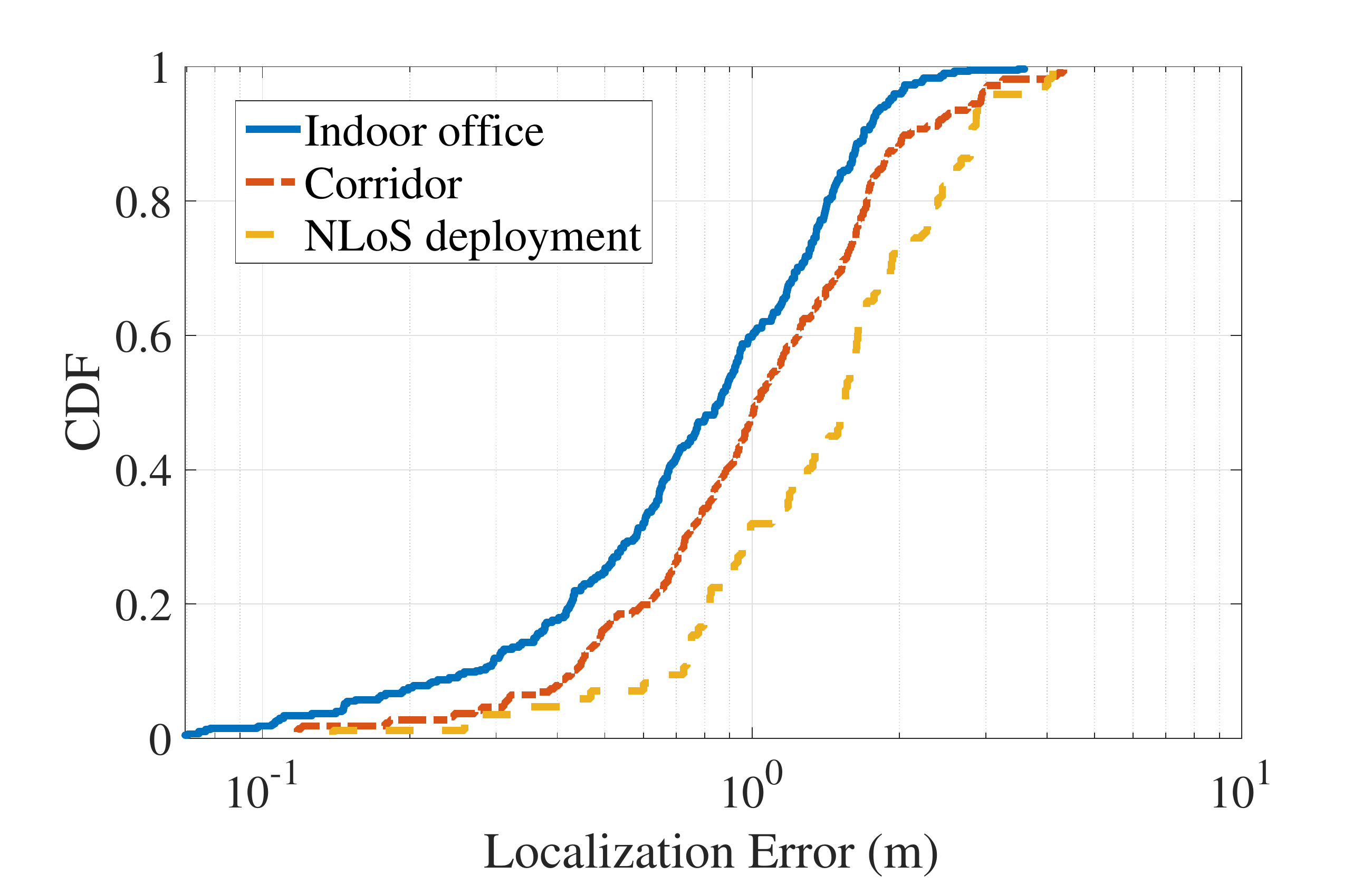}
		\caption{Impact of environment.} 
		\label{fig:overallper0723}
	\end{minipage} \hspace{0.2cm}
	\begin{minipage}[t]{0.45\textwidth}\centering
		\includegraphics[width=1\textwidth]{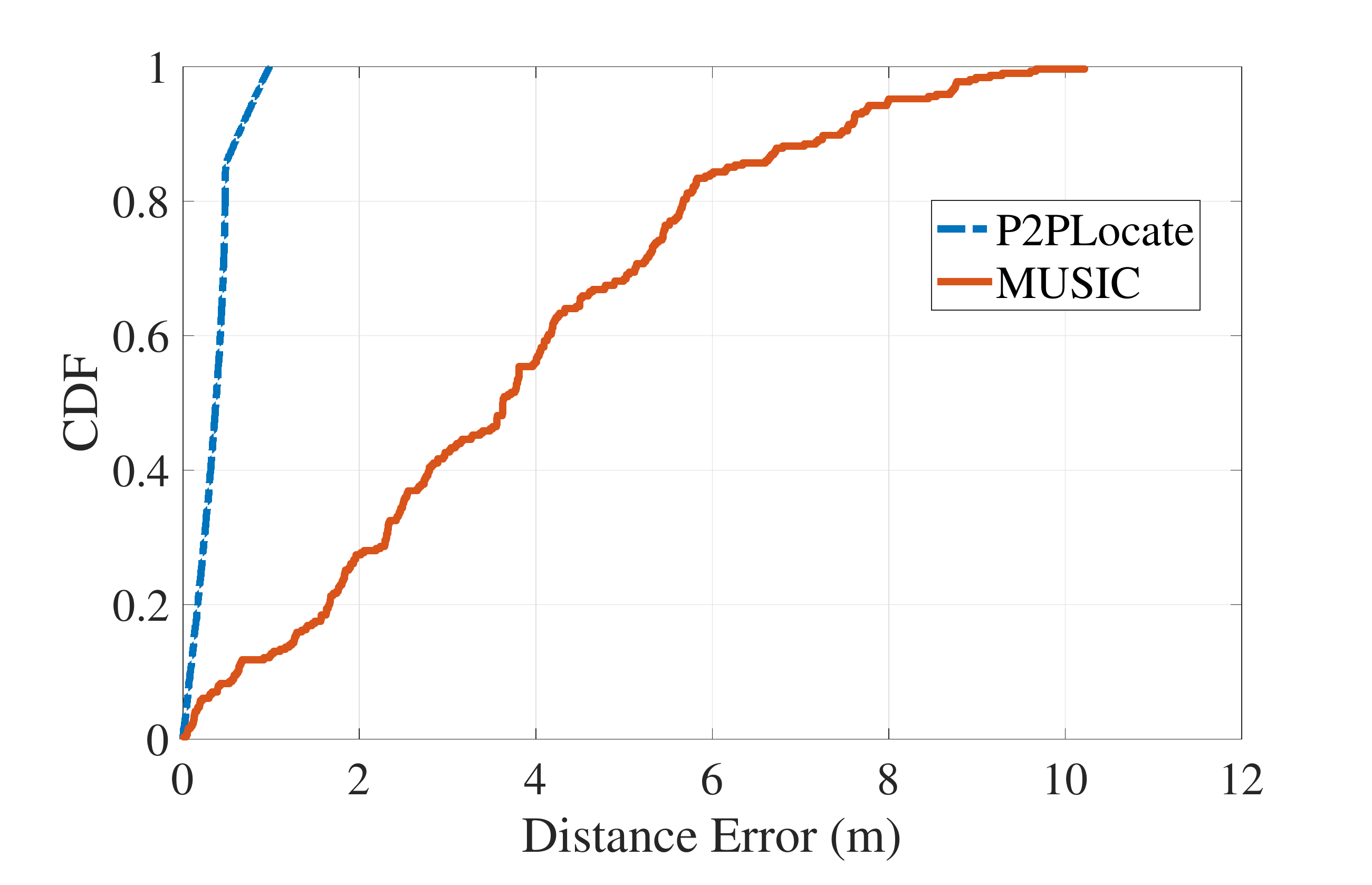}
		\caption{Range estimation accuracy.} 
		\label{fig:distanceError}
	\end{minipage}\hspace{0.2cm}

\end{figure*}

\noindent\textbf{Moving target.} Fig.~\ref{fig:distanceerror_motion} shows the localization results when both the localization end and the target move. \sys\ achieves a median localization accuracy of 0.98~m, while IMU-assisted solution, SpotFi and 2D mD-Track achieve 33~m, 0.82~m and 0.69~m accuracy, respectively. As seen in the figure, \sys\ achieves similar results compared to the case that the target remains static, while the IMU-assisted solution suffers larger errors. 

The reasons for similar performance for both moving and static cases in
\sys\ are that (i) \sys\ captures the relative movement of the transmitter
and the receiver, and (ii) a mobility-resilient direction estimation
algorithm is proposed to eliminate the effect of target's movement.
IMU-assisted solution fails in localizing the target accurately because the
onboard IMU sensors can only capture the absolute movement of the
localization end instead of the relative movement. Furthermore, the IMU
sensors suffer from large accumulative errors.

\subsection{Deep Dive into \sys}
The previous results show the overall accuracy of \sys\ for localization. Here, we zoom in on the details.

\subsubsection{Doppler Arithmetic Sign Estimation Accuracy of Doppler Shift}
First, we evaluate the robustness of Doppler direction estimation. \sys\
proposes a two-step estimation algorithm to estimate both the absolute value
and arithmetic sign of the Doppler shift, which are further used for both
range and direction estimation. We use an iRobot Create robot to carry a
transmitter and move close or away the receiver from different
directions with varying speed. As shown in Fig.~\ref{fig:dopplerdirection},
\sys\ yields an average estimation accuracy of 98.6$\%$ under different
moving speeds. The results demonstrate that \sys\ can identify whether the
parent is moving close or away the lost kid accurately. Overall,
\sys\ demonstrates the ability to estimate fine-grained range and direction
information, and thus the final position.

\subsubsection{Impact of Environments}
We also evaluate the localization accuracy in corridors and None Line-of-Sight (NLoS) environments. To ensure NLoS
deployment, the experiments are conducted in locations where the direct path is blocked by obstacles, such
as thick walls, wooden furniture, doors and windows. Specifically, the experiments of NLoS deployment are
conducted in laboratory 1 and office area as illustrated in Fig. 12. The receiver and the target are asked to be in
different rooms, or one in the room and the other in the corridor.  As shown in Fig.~\ref{fig:overallper0723}, the median localization accuracy of \sys\ is 1.02~m and 1.55~m in the corridor and NLoS deployments. The higher localization error rate in these two deployments is due to the nature of the backscatter which is sensitive to the signal power. However, the results are still robust due to our algorithms of backscatter CSI separation. 

\subsubsection{Range Estimation Accuracy}
To fully understand \sys's accuracy in range estimation, we conduct experiments in corridors with different lengths (from 10~m to 70~m). We control the receiver and the co-located backscatter tag to move along a straight line towards the transmitter at a speed of 1~m/s. The signals are transmitted and received at 5~GHz frequency band with 40~MHz bandwidth. We use the traditional MUltiple SIgnal Classification (MUSIC) algorithm~\cite{xiong2015tonetrack} that requires multiple CSI measurements across packets for comparison. As shown in Fig.~\ref{fig:distanceError}, \sys\ achieves a median accuracy of 0.37~m compared to 3.63~m by only using MUSIC algorithm. With 40~MHz signal bandwidth, it achieves a raw distance resolution of 7.5 meters. The better accuracy of \sys\ comes from the fusion of ToF and Doppler, which smooths and refines the range estimates.

\subsubsection{Angle Estimation Accuracy}
\begin{figure}
	%\begin{figure}[ht]
	\centering
	\subfigure[Angle estimation result.]{\label{fig19_1} 
		\includegraphics[width = 0.45\textwidth]{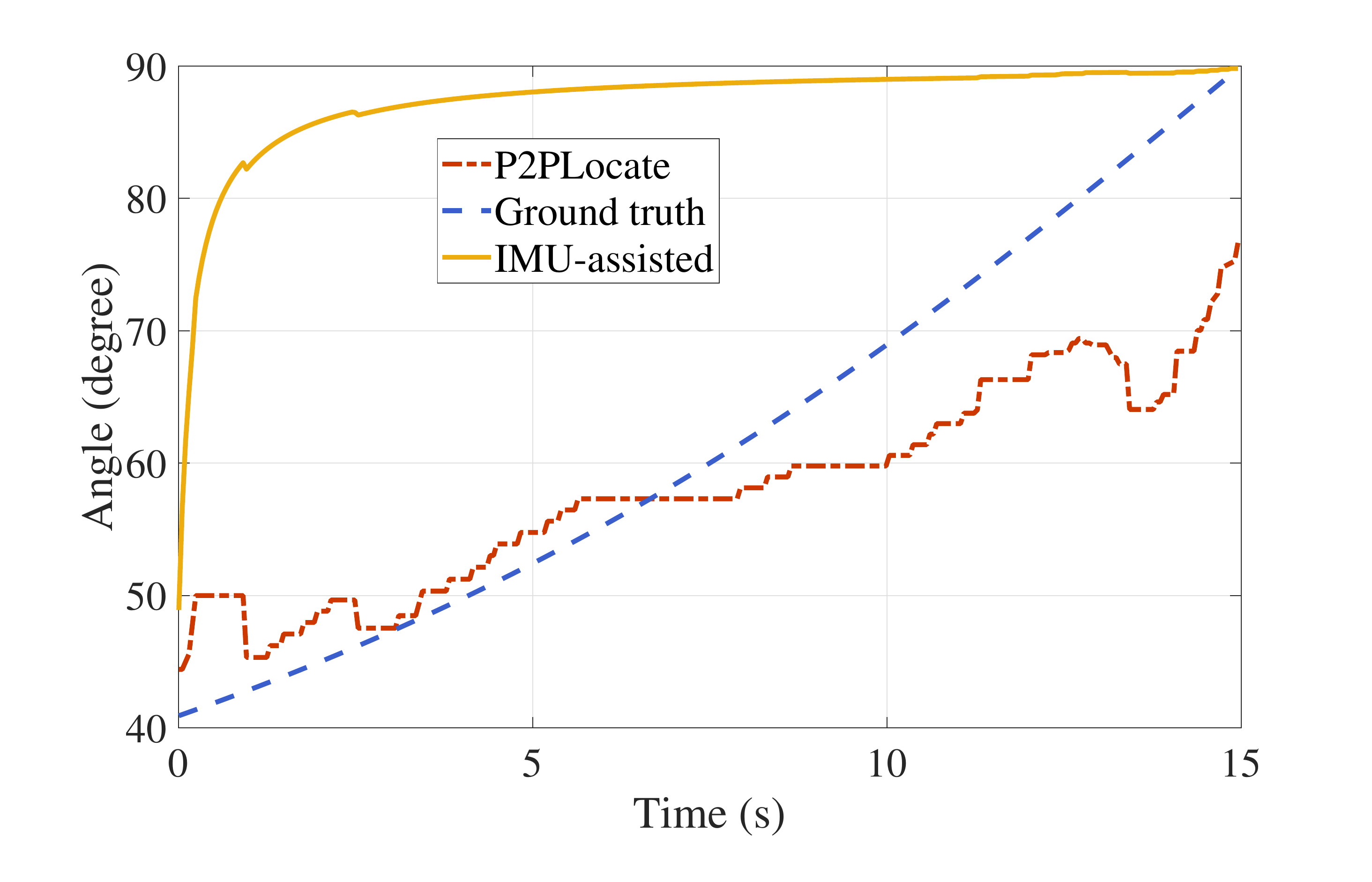}} 
	%\hspace{0.1in}
	\subfigure[Angle estimation accuracy.] {\includegraphics[width = 0.45\textwidth]{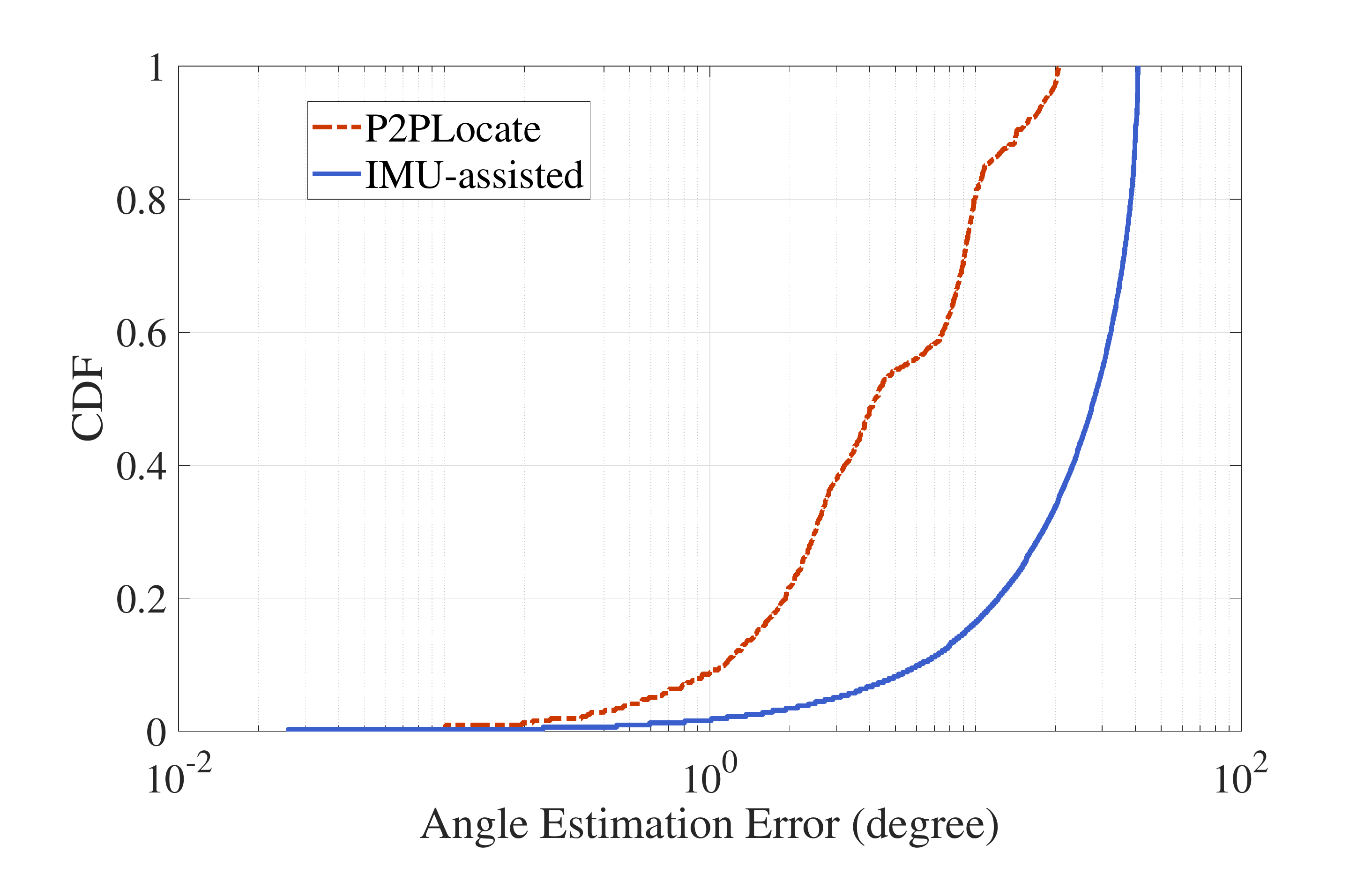}}
	%\hspace{1.7ex}
	\caption{Angle estimation.}
	\label{fig:AngleError}
	%\end{figure}
\end{figure}

We also evaluate the angle estimation accuracy of \sys. The experiments are
conducted in a hallway by using an iRobot Create robot to control the
receiver and the co-located backscatter tag to move in predetermined
trajectories, including circles, rectangles, and curves.
Fig.~\ref{fig:AngleError}(a) shows the angle estimation results. From the results, we observe that \sys\ estimates the angle accurately while the angle estimates by using IMU are highly unreliable. Fig.~\ref{fig:AngleError}(b) demonstrates that \sys\ achieves a median
accuracy of 4.2~degrees, while the IMU-assisted solution achieves a median
accuracy of 28.1~degrees. In \sys, the angle of the target is computed by
dividing the moving speed by radial speed of the receiver relative to the target. \sys\ achieves better performance due to its superior capability in moving speed estimation. While the IMU-assisted method suffers large accumulative error, leading to large error in speed estimation, and thus large error in angle estimation.

\subsubsection{Impact of Moving Speed} 

As stated in Section ~\ref{DirectionEst}, the direction of the target device
is estimated by computing the moving speed and radial speed. So, we
investigate the impact of moving speed on \sys\ by varying it from 0.5~m/s
to 3~m/s with an interval of 0.5~m/s by using an iRobot Create robot.
Fig.~\ref{fig:varyingspeed} plots the localization results including both
median accuracy and standard deviation. The results demonstrate \sys\
achieves a median localization accuracy of around 0.88~m, and the accuracy
improves with moving speed. The reason is that larger moving speed leads to larger DFS, which may facilitate the range estimation.

\subsubsection{Impact of Distance} \label{DistanceImpact}

\begin{figure*}[t]
	\centering 
	\begin{minipage}[t]{0.45\textwidth}\centering
		\includegraphics[width=1\textwidth]{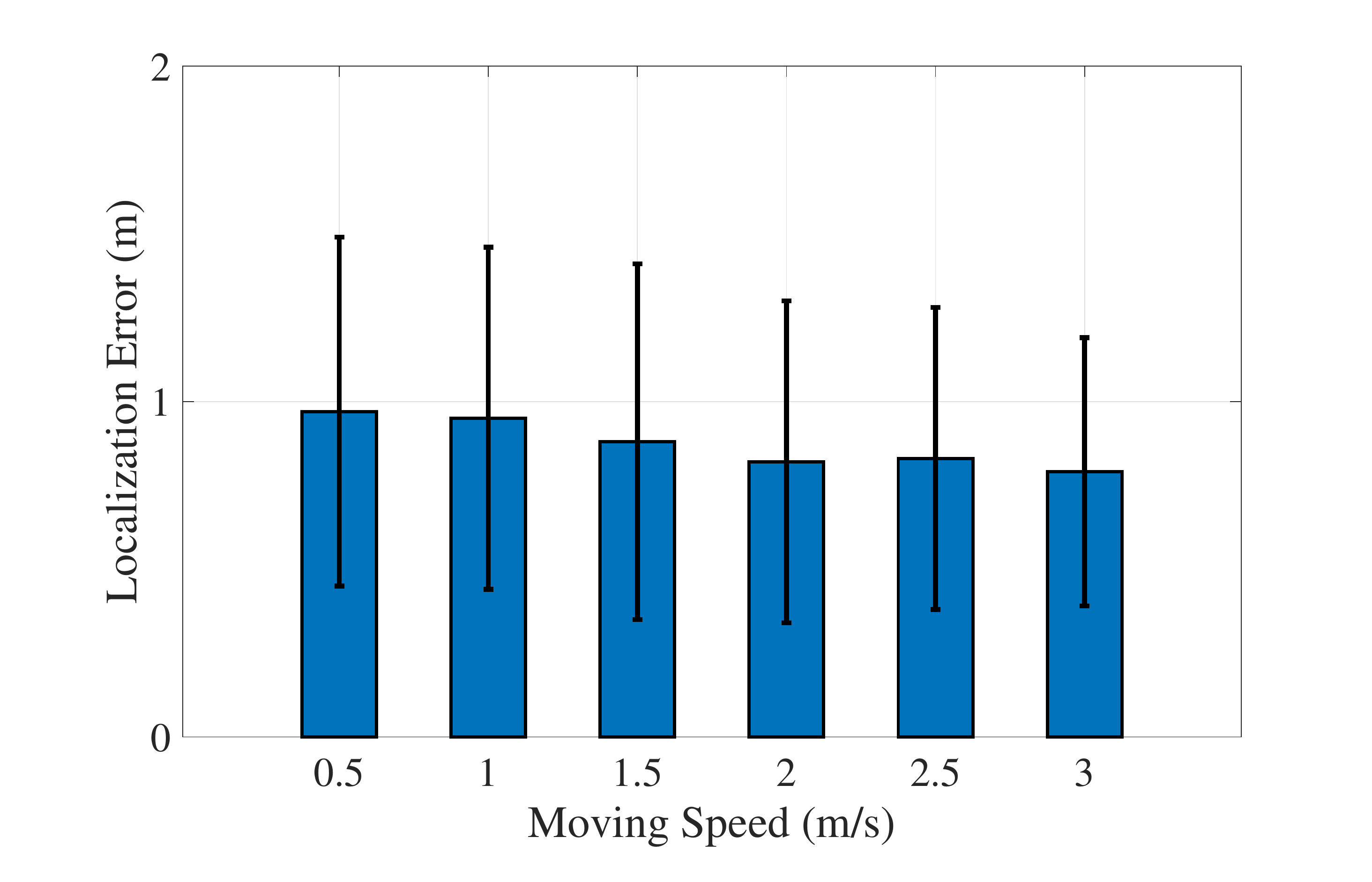}
		\caption{Impact of moving speed.}
		\label{fig:varyingspeed}
	\end{minipage} \hspace{0.2cm}
	\begin{minipage}[t]{0.45\textwidth}\centering
		\includegraphics[width=1\textwidth]{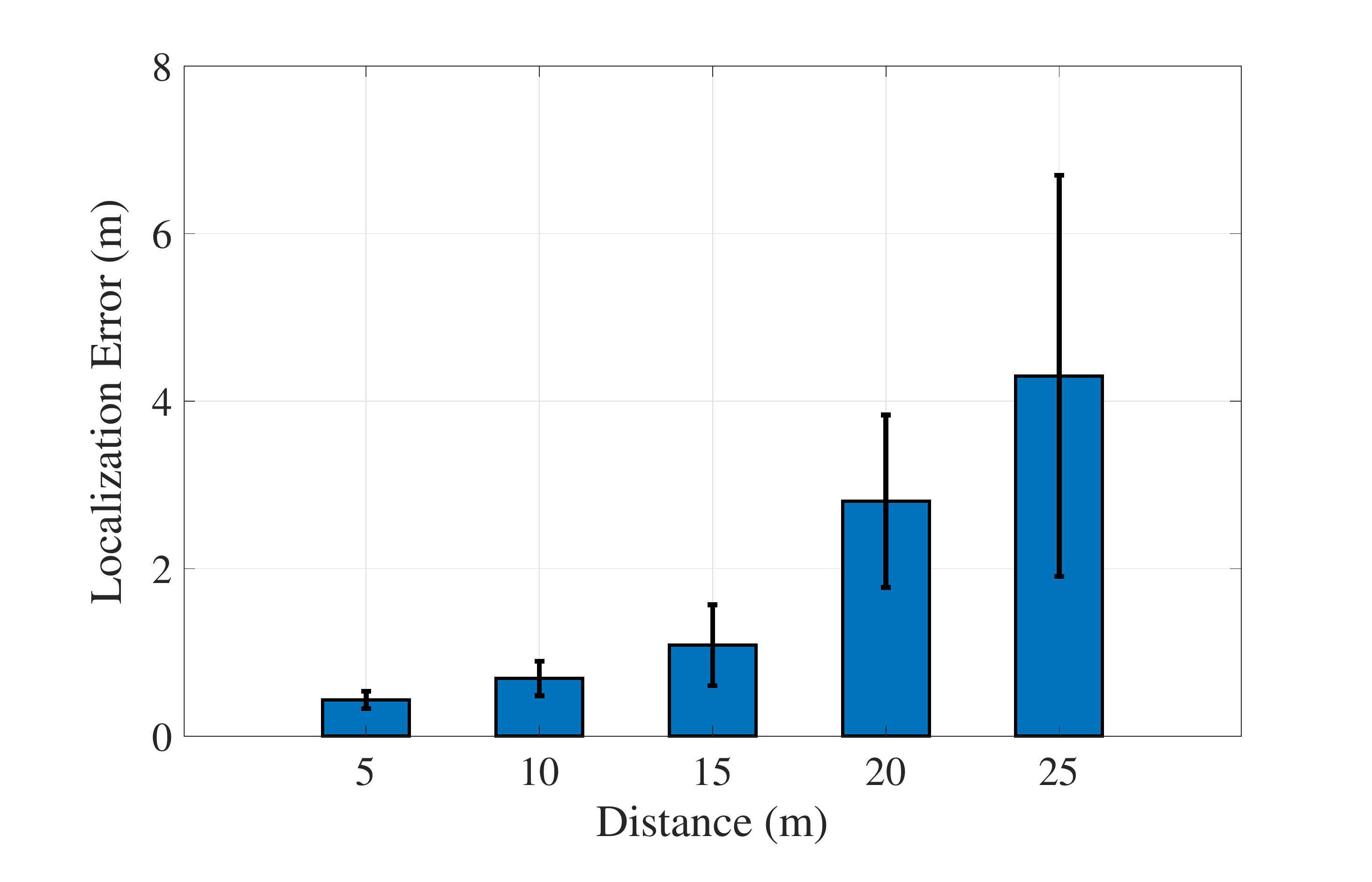}
		\caption{Impact of distance.}
		\label{fig:errordistance}
	\end{minipage} 
\end{figure*}

Next, we evaluate the impact of the distance between the transmitter and the receiver on localization accuracy. We conduct experiments by varying distance from 5~m to 25~m. Fig.~\ref{fig:errordistance} plots the median and standard deviation of localization error against the distance between the transceivers. The median localization error is initially around 0.43~m and increases to at most 4.3~m at 25~meters. In any localization system, the error scales with the distance between the transceivers. The reasons are as follows. First, the signal-to-noise ratio reduces at further distances, which causes the signals reflected by the backscatter tag to become too weak for robust CSI extraction. Second, with the same angle estimation error, larger distance leads to larger localization errors~\cite{kotaru2015spotfi}. It is worth noting that these errors are small enough to help patients find their pill bottles or find target good in a messy warehouse.

\subsubsection{Impact of Channel Bandwidth}
\begin{figure}
	\centering
	\includegraphics[width=0.5\linewidth]{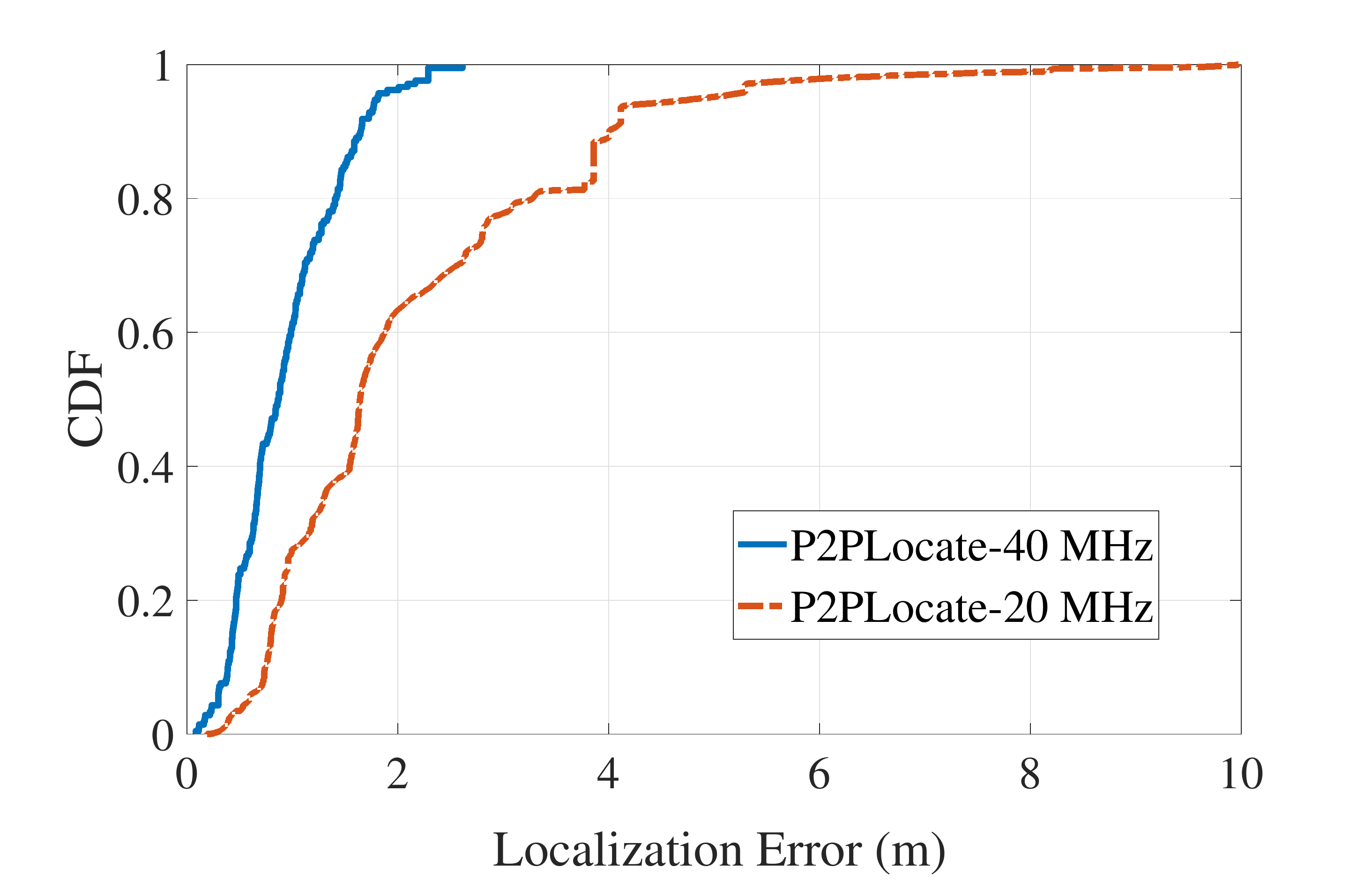}
	\caption{Impact of signal bandwidth.}
	\label{fig:bandwidth}
\end{figure}

Recall that the accuracy of ToF estimation directly depends on signal bandwidth. Therefore, we also evaluate the performance of \sys\ over 20~MHz signal bandwidth. As showed in Fig.~\ref{fig:bandwidth}, the localization accuracy of \sys\ reduces over 20~MHz signal bandwidth compared with 40~MHz bandwidth, which is similar to all ToF-based solutions. Yet, it still achieves a median localization accuracy of 1.64~m. This is because we estimate the range by using not only ToF estimates but also Doppler shift, which refines the range estimation and contributes the most to the final stable estimate. In addition, narrower signal bandwidth results in a larger coverage area of \sys\ since it leads to a longer communication range of the backscatter tags~\cite{xu2018practical}.

%\begin{figure}
%	\centering
%	\includegraphics[width=0.45\linewidth]{figures/varyingSpeed1}
%	\caption{Impact of moving speed.}
%	\label{fig:varyingspeed}
%\end{figure}
%
%
%\begin{figure}[t]
%	\centering
%	\includegraphics[width=0.45\linewidth]{figures/error_distance1} \vspace{-0.2cm}
%	\caption{Impact of distance.}
%	\label{fig:errordistance}
%\end{figure}

\section{Related Work}\label{sec:related}

\textbf{Indoor localization.} Numerous efforts have been devoted to developing indoor localization technologies during the past two   decades~\cite{sen2012you,ArrayTrack,sugasaki2017robust,chen2017taming,khan2009diland,zhang2019rf}. Most  existing works rely on pre-deployed indoor infrastructures to estimate different channel parameters of the direct-path signal for localization, such as ToF~\cite{sen2013avoiding,vasisht2016decimeter}, AoA~\cite{ArrayTrack} or their fusion~\cite{kotaru2015spotfi}. These approaches push the localization accuracy to decimeter level by using antenna arrays ~\cite{kotaru2015spotfi,ArrayTrack} or ultra-wide channel bandwidth ~\cite{xie2018precise}. However, They typically require coordination between multiple APs and are not available in our target use cases. In contrast, \sys\ enables a single receiver to localize another device and achieves decimeter-level accuracy.

%WiTag~\cite{kotaru2017localizing} as mentioned in the comment as [1] leverages multiple three-antenna APs to localize low-power backscatter tags which is not applicable in peer-to-peer localization scenario. ppNav~\cite{yin2017peer} as mentioned in the comment as [2] enables peer-to-peer navigation using smartphones by utilizing wireless fingerprints, which requires multiple APs. In contrast, \sys\ requires no infrastructures and only a single receiver. Other peer-to-peer localization systems that use only a single receiver however require multiple antennas or ultra-wide signal bandwidth. Specifically, SAIL~\cite{mariakakis2014sail} achieves a median accuracy of 2.3~m by coupling AoA measured by multiple antennas and ToF measured by a build-in 88~MHz clock of the WiFi card. Chronos~\cite{vasisht2016decimeter} combines all available WiFi frequency bands to achieve decimeter-level accuracy, which may affects ongoing WiFi communications. A recent proposal has demonstrated decimeter-level accuracy with a single AP and single-channel measurement by utilizing the multipath reflections~\cite{soltanaghaei2018multipath}. However, it requires a 3-element antenna array on either the AP side or the target side. In contrast, \sys\ enables peer-to-peer localization by only relying on the low-profile on-body devices, and achieves sub-meter level accuracy, with only a single antenna on both the transmitter and receiver.

There have been some recent advances in developing peer-to-peer localization systems. SAIL~\cite{mariakakis2014sail} achieves a median accuracy of 2.3~m by coupling AoA measured by multiple antennas and ToF measured by a build-in 88~MHz clock of the WiFi card. Chronos~\cite{vasisht2016decimeter} combines all available WiFi frequency bands to achieve decimeter-level accuracy, which may affect ongoing WiFi communications. A recent proposal has demonstrated decimeter-level accuracy with a single AP and single-channel measurement by utilizing the multipath reflections~\cite{soltanaghaei2018multipath}. However, it requires a 3-element antenna array on either the AP side or the target side. In contrast, \sys\ enables peer-to-peer localization by only relying on the low-profile on-body devices, and achieves sub-meter level accuracy, with only a single antenna on both the transmitter and receiver and a single channel measurement.

Other works employ synthetic aperture radar sensing algorithms to form a virtual antenna
array~\cite{wang2013dude,kumar2014accurate}, which can be used to identify the
spatial direction of incoming wireless signals. Thus, these proposals can
achieve fine-grained localization without hardware modification. However, they
need to precisely estimate the location of each virtual antenna and cannot
locate moving targets. In contrast, \sys\ does not need precise antenna
position information and is resilient with mobile target. Sen et
al.~\cite{sen2013avoiding} fuse motion sensor data with wireless signals for
localization, which however suffers large accumulative errors. The IEEE 802.11-2016~\cite{80211} standardized a Fine Time Measurement (FTM) for range estimation, which offers meter-level accuracy. However, it cannot obtain range estimation corresponding to the backscatter and the localization device simultaneously. Many other
modalities can also be used to localize objects, such as visible
light~\cite{zhang2016litell,zhang2017pulsar}, acoustic
sensing~\cite{mao2016cat,yun2015turning}, electromagnetic
field~\cite{lu2018simultaneous}, etc. These technologies achieve good
performance but require dedicated hardware or are vulnerable to dynamic ambient
contexts, making them unsuitable in our targeted applications. 

\noindent\textbf{Indoor Human Tracking.} Indoor human tracking has attracted a
lot of attention over the past a few years~\cite{li2016dynamic,vasisht2018duet,wang2016gait,mTrack}. Some recent proposals have demonstrated sub-meter level accuracy by using the reflected signals from human bodies. Wi-Vi~\cite{adib2013see} captures reflections from multiple moving objects behind a wall to track them and even identify simple gestures. However, it can only track relative movements and requires dedicated devices. WiDeo~\cite{joshi2015wideo} achieves fine-grained motion recognition by using software-defined radios, which limits its applications for commodity devices. Some other solutions, such as IndoTrack~\cite{li2017indotrack}, Widar2.0~\cite{qian2018widar2}, and mD-Track~\cite{xie2019md}, leverages Doppler shift for fine-grained human or motion tracking by using commodity devices. However, they require multiple antennas for estimating the direction information of Doppler shift. In comparison, instead of multiple antennas, \sys\ leverages the human movement along with the CSI power to estimate both the absolute value and arithmetic sign of Doppler shift, which is fully compatible with single-antenna devices.

\noindent\textbf{Backscatter communications.} Backscatter communication is considered to
be one of the most prominent solutions to provide ubiquitous communication
capabilities for low-power Internet of Things. In recent years, there are many improvements in
throughput and communication range of backscatter systems
~\cite{bharadia2015backfi,kellogg2016passive,zhang2016hitchhike,liu2013ambient,song2019reliable}.
Liu et al.~\cite{liu2013ambient} enable communication between two battery-free
devices by leveraging ambient TV signals. Other
systems~\cite{zhang2016hitchhike,bharadia2015backfi,kellogg2014wi} backscatter
information on top of WiFi signals and use off-the-shelf WiFi devices as
the transceiver. Based on the backscatter communication paradigm, many works focus
on using backscatter signals for activity recognition, IoT security or
localization~\cite{ryoo2018barnet,xiao2018motion,kotaru2017localizing,huang2019lightweight}. Ryoo et
al.~\cite{ryoo2018barnet} and Xiao et al.~\cite{xiao2018motion} use backscatter
signals to enable activity recognition when multiple users perform certain
motions simultaneously. WiTag~\cite{kotaru2017localizing} uses backscatter
signals and more than three APs to localize backscatter tags based on AoA
measurements. In contrast, \sys\ leverages the multipath variations offered by the backscatter modulation to identify the direction of the target and thus localize the target accurately.

\section{Discussion}\label{sec:discussion}

In this work, we focus on developing an indoor peer-to-peer localization system, which achieves decimeter-level accuracy with single-antenna transceivers. We briefly discuss some limitations and practical issues that have not been directly addressed in this paper.

\subsection{Absence of Direct Path}

In our design, we assume that there exists a direct path between the receiver and the target, which is a common assumption shared by most localization systems~\cite{soltanaghaei2018multipath, xie2018precise, luo20193d}. The evaluation demonstrates that \sys\ works well in both real-world LoS and NLoS deployments. However, in the extreme case that the direct path is completely blocked, it will fail to localize the target reliably. In the future, we plan to use the information of the reflected paths from as many dimensions as possible, e.g., ToF, AoA and angle of departure (AoD), to localize the target while there exists no direct path.

%	One potential solution to address this is to leverage the multipath reflections to simultaneously localize the target and mapping the environments.
%	
%	P2PLocate relies on the existence of the direct-path propagation between the transmitter and the receiver, which is a common limitation shared by most of WiFi localization systems. However, the signal traveled along the direct path could be completely blocked due to the presence of serious occlusions. This could leads to large error in location estimates. One future direction to address this challenge could be leverage the multiple-dimension information corresponding to the reflected signals to simultaneously localize the target and mapping the environments.}
%
%{\color{blue} \textit{Requirement of one backscatter tag.} P2PLocate requires one backscatter tag to achieve peer-to-peer localization within decimeter-level accuracy. However, more general solutions may be required for wide adoption. A potential solution is to enable peer-to-peer localization by integrating wireless and vision information which is universally available in most mobile devices.}

\subsection{Localization Range}
Due to the requirement of robust backscatter CSI separation, the system's localization range in our experiments is limited up to 25~m. It is a common range used in localization systems~\cite{vasisht2016decimeter} and backscatter communications~\cite{zhang2016hitchhike}. \sys\ can work well in most applications mentioned in Section.~\ref{sec:Introduction} at this range, e.g., help patients find their pill bottles. To cover a larger space, we can improve the reliability of backscatter CSI separation at further distances by leveraging recent advances in tunnel diodes based backscatter designs~\cite{varshney2019tunnelscatter}. Since tunnel diodes only amplify the reflected signal, we believe \sys\ would also work when using tunnel diodes based backscatter tags.

%The main reason is that the signals reflected by the backscatter tag becomes too weak for robust detection at furtherdistance. In the future, we can use tunnel scatter to amplify the reflected signal and significantly enhancethe localization range of P2PLocate. Future, we plan to deploy P2PLocate in more challenging and larger space, such as shopping malls and warehouses with dense crowds, to demonstrate large-scale applications.}

\subsection{Impact of Obstacles}

In any localization system~\cite{soltanaghaei2018multipath, xie2018precise, luo20193d}, obstacles will affect the performance of \sys. In particular, more obstacles may lead to more serious multipath and affect the accuracy of ToF estimation. Besides, multipath reflections from moving obstacles may cause some adversary effect on Doppler estimation. In our target application scenarios, in addition to the direct path, there are usually 1 to 4 dominant paths reflected by the obstacles~\cite{kotaru2015spotfi, gong2018sifi, ArrayTrack}. \sys's current design works well in such environments. To extent \sys\ in the extreme environments full of obstacles, we may use wider signal bandwidth, e.g., 160~MHz in 802.11ac standard, to enhance the ability of resolving reflections from multiple obstacles. Besides, we may use a larger observation window in Doppler estimation to improve the resolution of resolving Doppler effects caused by mobile obstacles.

\subsection{WiFi and Backscatter Interference}
In our experiments, we test \sys\ in the environment where there are tens of WiFi devices within the interference range. As our system builds on commodity WiFi devices, it is resilient to the interference from other WiFi devices as standard WiFi systems, which uses Carrier Sense Multiple Access with Collision Avoidance (CSMA/CA) to avoid interference. To enhance the performance under interference-rich environments, our future implementations of \sys\ can integrates some recent advances in WiFi interference cancellation technologies~\cite{sen2011csma}. In addition, the backscatter interference may degrade the performance of \sys\, which can be alleviated by employing the MAC layer protocol proposed in~\cite{zhang2017freerider}.

\section{Conclusion}\label{sec:conclusion}

In this paper, we present the design, implementation, and evaluation of \sys, a peer-to-peer localization technique that leverages on-body devices to enable a single-antenna WiFi device to locate another WiFi device in indoor environments. Real-world evaluations demonstrate that \sys\ achieves decimeter-level accuracy with only a single channel and a single antenna in both the transmitter and the receiver and without the support of pre-deployed infrastructures or pre-training, allowing its wide-scale adoption due to the plug-and-play manner. By doing so, \sys\ opens up WiFi-based localization to new applications with size-constrained devices and infrastructure-free environments. %We believe AnchorScatter can also benefit current indoor RF localization and navigation services. 

\begin{acks}
	This work was supported in part by the National Key R\&D Program of China under Grant 2019YFB180003400, Young Elite Scientists Sponsorship Program by CAST under Grant 2018QNRC001, and National Science Foundation of China with Grant 91738202.
\end{acks}

\bibliographystyle{ACM-Reference-Format}

\bibliography{sample-sigconf}

\end{document}